\hsize=14 cm \vsize=20.8 cm \tolerance=400
\hoffset=2cm
\font\trm = cmr10 scaled \magstep3
\font\srm = cmr10 scaled \magstep2

\voffset=2cm
\scriptscriptfont0 =\scriptfont0
\scriptscriptfont1 =\scriptfont1

\def\d{\partial}
\def\dh{\mathop{\vphantom{\odot}\hbox{$\partial$}}}
\def\dl{\dh^\leftrightarrow}
\def\sqr#1#2{{\vcenter{\vbox{\hrule height.#2pt\hbox{\vrule width.#2pt 
height#1pt \kern#1pt \vrule width.#2pt}\hrule height.#2pt}}}}
\def\w{\mathchoice\sqr45\sqr45\sqr{2.1}3\sqr{1.5}3\,}

\def\psq{{\overline{\psi}}}

\def\=d{\,{\buildrel\rm def\over =}\,}

\def\i3p{\p32\int d^3p}

\def\As{A\hbox to 1pt{\hss /}}
\def\np4{\int d^4p_1\cdots d^4p_{n-1}\, }

\def\nx4{\int d^4x_1\ldots d^4x_n\, }

\def\kon#1#2{\vbox{\halign{##&&##\cr
\lower4pt\hbox{$\scriptscriptstyle\vert$}\hrulefill &
\hrulefill\lower4pt\hbox{$\scriptscriptstyle\vert$}\cr $#1$&
$#2$\cr}}}

\def\konv#1#2#3{\hbox{\vrule height12pt depth-1pt}
\vbox{\hrule height12pt width#1cm depth-11.6pt}
\hbox{\vrule height6.5pt depth-0.5pt}
\vbox{\hrule height11pt width#2cm depth-10.6pt\kern5pt
      \hrule height6.5pt width#2cm depth-6.1pt}
\hbox{\vrule height12pt depth-1pt}
\vbox{\hrule height6.5pt width#3cm depth-6.1pt}
\hbox{\vrule height6.5pt depth-0.5pt}}
\def\konu#1#2#3{\hbox{\vrule height12pt depth-1pt}
\vbox{\hrule height1pt width#1cm depth-0.6pt}
\hbox{\vrule height12pt depth-6.5pt}
\vbox{\hrule height6pt width#2cm depth-5.6pt\kern5pt
      \hrule height1pt width#2cm depth-0.6pt}
\hbox{\vrule height12pt depth-6.5pt}
\vbox{\hrule height1pt width#3cm depth-0.6pt}
\hbox{\vrule height12pt depth-1pt}}

\def\konw#1#2#3{\hbox{\vrule height12pt depth-1pt}
\vbox{\hrule height12pt width#1cm depth-11.6pt}
\hbox{\vrule height6.5pt depth-0.5pt}
\vbox{\hrule height12pt width#2cm depth-11.6pt \kern5pt
      \hrule height6.5pt width#2cm depth-6.1pt}
\hbox{\vrule height6.5pt depth-0.5pt}
\vbox{\hrule height12pt width#3cm depth-11.6pt}
\hbox{\vrule height12pt depth-1pt}}

\def\i{{\rm int}}

\def\m3{{\mu_1\mu_2\mu_3}}

\def\p{{(+)}}

\nopagenumbers
\vbox to 1.5cm{ }
\centerline{Preprint: ZU-TH-30/95, hep-th/9606105}
\vbox to 1.5cm{ }
\centerline{\trm Slavnov-Taylor Identities from the Causal Point of View }
\vskip 2cm
\centerline{\srm Michael D\"utsch } \vskip 0.5cm
\centerline{\it
Institut f\"ur Theoretische Physik der Universit\"at Z\"urich}
\centerline{\it Winterthurerstr. 190, CH-8057 Z\"urich, Switzerland}\vskip 3cm
{\bf Abstract.} -We continue the investigation of quantized
Yang-Mills theories coupled to matter fields in the framework of causal
perturbation
theory which goes back to Epstein and Glaser.
In this approach gauge invariance is expressed by a simple commutator relation
for the
S-matrix and the corresponding gauge transformations are simple transformations
of the free fields only. In spite of this simplicity, gauge invariance implies
the usual
Slavnov-Taylor identities. The main purpose of this paper is to prove the
latter statement.
Since the Slavnov-Taylor identities are formulated in terms of Green's
functions, we
investigate the agreement of two perturbative definitions of Green's functions,
namely of Epstein and Glaser's definition with the Gell-Mann Low series.
chromodynamics.
\vfill\eject
\pageno=1
\headline={\tenrm\ifodd\pageno\hss\folio\else\folio\hss\fi}
\footline={\hss}
{\trm 1. Introduction}
\vskip 1cm
{\it (a) The Model}
\vskip 0.5cm
In a recent series of papers [1,2,3,4,5] non-abelian gauge invariance has been
studied
in the framework of causal perturbation theory. This approach, which goes back
to
Epstein and Glaser [6], has the merit that one works exclusively with free
fields, which
are mathematically well-defined, and performs only justified operations
with them. Consequently, the gauge transformations are transformations of the
free
fields only.

In causal perturbation theory one makes
an ansatz for the S-matrix as a formal power series in the coupling constant
$$S(g)=1+\sum_{n=1}^{\infty}{1\over n!}\int d^4x_1...
d^4x_n\,T_n(x_1,...,x_n)g(x_1)...g(x_n).\eqno(1.1)$$
The test function $g\in{\cal S}({\bf R}^4)$ switches the interaction and
$T_n(x_1,...,x_n)$ is an operator-valued distribution. The $T_n$'s are
constructed
inductively from the given first order
$$T_1(x)=T_1^A(x)+T_1^u(x)+T_1^\psi (x),\eqno(1.2)$$
with
$$T^A_1(x)\=d {ig\over 2}f_{abc}:A_{\mu a}(x)A_{\nu
b}(x)F^{\nu\mu}_c(x):,\eqno(1.3)$$
$$T^u_1(x)\=d -igf_{abc}:A_{\mu a}(x)u_b(x)\d^\mu\tilde u_c(x):,\eqno(1.4)$$
$$T_1^\psi (x)\=d i\,j_{\mu a}(x)A_a^\mu(x),\eqno(1.5)$$
where the matter current $j_{\mu a}$ is defined by
$$j_{\mu a}(x)\=d {g\over
2}:\psq_\alpha(x)\gamma_\mu(\lambda_a)_{\alpha\beta}\psi_\beta(x):.
\eqno(1.6)$$
Herein, $g$ is the coupling constant, $f_{abc}$ are the structure constants of
the group
SU(N) and ${-i\over 2}\lambda_a,\>a=1,...,N^2-1$ denote the generators of the
fundamental
representation of SU(N).
The gauge potentials $A^\mu_a,\>F_a^{\mu\nu}\=d \d^\mu A_a^\nu -\d^\nu A_a^\mu
$,
and the ghost fields $u_a,\>\tilde u_a$ are massless and fulfil the wave
equation.
The matter fields $\psi_\alpha$ and $\psq_\alpha\=d \psi_\alpha^+\gamma^0$
satisfy
the free Dirac equation with a colour independent mass $m\geq 0$ [5].
Therefore,
the matter current is conserved
$$\d^\mu j_{\mu a}(x)=0.\eqno(1.7)$$

{\it Gauge invariance} means roughly speaking
that the commutator of the $T_n$-distributions with the gauge charge
$$Q\=d \int_{t=const.} d^3x\,(\d_{\nu}A_a^{\nu}{\dl}_0u_a)\eqno(1.8)$$
is a (sum of) divergence(s). In first order this holds true
$$[Q,T^A_1(x)+T^u_1(x)]=i\d_\nu
(T^{A\nu}_{1/1}(x)+T^{u\nu}_{1/1}(x)),\eqno(1.9)$$
where
$$T^{A\nu}_{1/1}(x)\=d igf_{abc}:A_{\mu a}(x)u_b(x)F_c^{\nu
\mu}(x):,\eqno(1.10)$$
$$T^{u\nu}_{1/1}(x)\=d -{ig\over 2}f_{abc}:u_a(x)u_b(x)\d^\nu\tilde
u_c(x):,\eqno(1.11)$$
and, by means of the current conservation (1.7),
$$[Q,T_1^\psi (x)]=i\d_\nu T^{\psi\nu}_{1/1}(x)\eqno(1.12)$$
with
$$T^{\psi\nu}_{1/1}(x)\=d i\,j_a^\nu (x)u_a(x).\eqno(1.13)$$
Note that $[Q,T^A_1]$ alone is not a divergence. In order to have gauge
invariance in first
order, we are forced to introduce the ghost coupling $T^u_1$ (1.4). We define
gauge
invariance in arbitrary order by
$$[Q,T_n(x_1,...,x_n)]=i\sum_{l=1}^n\d_\mu^{x_l}T^\mu_{n/l}(x_1,...,x_n).
\eqno(1.14)$$
The divergences on the r.h.s. of (1.14) are given by $n$-th order
$T$-distributions
from a different theory which contains, in addition to the usual
Yang-Mills couplings (1.2), a so-called Q-vertex, defined by
$T^\nu_{1/1}=T^{A\nu}_{1/1}
+T^{u\nu}_{1/1}+T^{\psi\nu}_{1/1}$ (1.10), (1.11), (1.13). (See ref. [2] for
more details.)
Gauge invariance (1.14) implies the invariance of the S-matrix (1.1) with
respect to simple gauge
transformations of the {\it free} fields [5]. These transformations are the
{\it free field
version of the famous BRS-transformations} [7]. Moreover, {\it unitarity on the
physical
subspace} can be proven by means of gauge invariance (1.14) [4].

As usual, the $T_n$'s are constructed in normally ordered form
$$T_{n(/l)}(x_1,...,x_n)=\sum_{\cal O}t^{(l)}_{\cal O}(x_1-x_n,...,x_{n-1}-x_n)
:{\cal O}(x_1,...,x_n):,\eqno(1.15)$$
where ${\cal O}(x_1,...,x_n)$ is a combination of free field operators and
$t^{(l)}_{\cal O}$
is a C-number distribution. The latter contains an {\it undetermined
but finite normalization polynomial}
$$t^{(l)}_{\cal O}(x_1-x_n,...)+\sum_{|a|=0}^{\omega ({\cal O})}
C_aD^a\delta^{(4(n-1))}
(x_1-x_n,...,x_{n-1}-x_n)\eqno(1.16)$$
of degree $\omega ({\cal O})$. One can prove [2,8]
$$\omega ({\cal O})=4-b-g-d-{3\over 2}f,\eqno(1.17)$$
where $b$ is the number of gauge bosons $(A,\,F)$,
$g$ the number of ghosts $(u,\,\d\tilde u)$, $d$ the number of derivatives
$(F,\,\d\tilde u)$ and
$f$ is the number of pairs $(\psq,\,\psi)$ in ${\cal O}$. The fact that
$\omega$ is
bounded in the order $n$ (here it is even independent on $n$)
is the {\it (re)normalizability} of this model.

The constants $C_a$ in (1.16) are restricted by symmetry properties, especially
by gauge invariance. The most important example is the second order tree
diagram
$b=4,\>g=f=d=0$, which has the normalization term
$$Cig^2:A_{\mu a}(x_1)A_{\nu
b}(x_1)A_d^\mu(x_2)A_e^\nu(x_2):f_{abc}f_{dec}\delta (x_1-x_2).
\eqno(1.18)$$
Gauge invariance (1.14) fixes the value of $C$ uniquely: $C=-{1\over 2}$ [1,5].
With this
value, (1.18) agrees with the usual {\it four-gluon coupling}. It propagates to
higher orders
in the inductive construction of the $T_n$'s (sect. 4.2 of [9]).
\vskip 0.5cm
{\it (b) The Aim of the Paper}
\vskip 0.5cm
The operator gauge invariance (1.14) can be expressed by the {\it
Cg-identities},
the C-number identities for gauge invariance, which imply (1.14). The
Cg-identities
have been proven by induction on the order $n$ in [2,3,4,5]. However, they
still
contain the Q-vertex, which is only a mathematical auxiliary tool to formulate
gauge invariance
(1.14). Our first goal will be {\it to eliminate the Q-vertex from the
Cg-identities}. This is done
in sect.2 by inserting the Cg-identities into each other and taking the limit
of vanishing {\it inner} momenta. We call the resulting equations reduced
Cg-identities.
Moreover, we study the (finite) renormalizations (1.16) preserving
the latter identities rsp. the original Cg-identities.

As mentioned above our gauge transformations in the perturbative S-matrix (1.1)
involve free fields
only. Especially a local transformation of the matter fields is excluded, since
$e^{i\Lambda (x)}\psi_\alpha (x)$ does not fulfil the free Dirac equation.
One may ask, therefore, whether (1.14) contains the full information of
non-abelian
gauge invariance. It is the aim of this paper to prove this, apart from the
independence on the gauge fixing, which is not considered here. (We always work
in Feynman gauge.) The usual gauge or BRS invariance is expressed by the famous
{\it Slavnov-Taylor identities} [10,11,12]. In sect.3 {\it we prove the latter
in
the framework of perturbation theory by means of our reduced Cg-identities}.
There is a interesting speciality concerning matter fields: For the
Cg-identities with a
pair $(\psq,\,\psi)$ among the external legs, the Q-vertex cannot be eliminated
completely. But
in this case also {\it Taylor [10] was forced to introduce the Q-vertex
$T_{1/1}^\psi$ (1.13)} to formulate his identities.

The Slavnov-Taylor identities are written in terms of connected or in terms
of one-particle irreducible Green's functions. The latter cannot be expressed
directly by our $t_{\cal O}$-distributions (1.15), since in higher orders the
$t_{\cal O}$'s contain one-particle reducible terms. We must go another way:
We work with the Gell-Mann Low series [13] for the connected Green's
functions.
Epstein and Glaser [6] have given another definition of Green's functions in
the framework
of causal perturbation theory. In the appendix we prove that their definition
agrees with the Gell-Mann Low expressions at least for massive theories.
\vskip 0.5cm
{\it (c) Notations}
\vskip 0.5cm
We recall our convention given in [2-5] of denoting numerical
distributions of {\it non-degenerate terms},
i.e. connected terms with at most one external field operator at each vertex.
The expression
$$t_{AB\ldots ab\ldots}^{\alpha
2}(x_2,x_3,\ldots):A_a(x_2)B_b(x_3)\ldots:\eqno(1.19)$$
means the sum of all connected diagrams with external field operators (legs)
$A_a(x_2)$,
$B_b(x_3)$..., $a$ and $b$ are colour indices.
(We only write the index $\tilde u$ instead of $\d\tilde u$
for an external field operator $\d_\nu\tilde u(x_i)$.) The subscripts $\alpha
2$ show
that this term belongs to $T^\alpha_{n/3}(x_1,x_2,x_3,\ldots)$ with
$Q$-vertex at the second argument ($=x_3$) of the numerical distribution $t$.
An immediate consequence of this notation is the relation
$$t_{AB\ldots ab\ldots}^{\alpha 1}(x_1,x_2,\cdots)=\pm t_{BA\ldots ba
\ldots}^{\alpha 2}(x_2,x_1,\cdots),\eqno(1.20)$$
where we have a minus sign, if $A, B$ are both ghost or both matter operators
and a
plus sign in all other cases.
The Lorentz indices of the two operators $A, B$ must
also be reversed. Note that the latter equation particularly holds, if $A$
and $B$ are the same field operators. Moreover, the sum over permutations of
the vertices
is present in $T_{n(/l)}$. For example,
$$T_n(x_1,...x_n)=\sum_{i,j,k,l=1}^n:\psq (x_i)t^{\mu\nu}_
{\psq\psi AA\,ab}(x_i,x_j,x_k,x_l,x_1...\bar x_i,\bar x_j,\bar x_k,\bar
x_l...,x_n)\cdot$$
$$\cdot\psi (x_j)A_{\mu a}(x_k)A_{\nu b}(x_l):+...+({\rm
disconnected\,\,terms}),\eqno(1.21)$$
where the coordinates with bar must be omitted and the sum runs only over the
region
$i\not=j,\>i\not=k,\>i\not=l,\>j\not=k,\>j\not=l$ and $k<l$. The latter
restriction is
not $k\not=l$, since the external field operators at the vertices $x_k$ and
$x_l$
agree. The matrix multiplication $:\psq t(...)\psi :$
concerns the spinor space and the space of the fundamental representation. We
emphasize
that in higher orders our $t$-distributions contain one-particle reducible
terms.

The {\it degenerate terms} have at least one vertex with two external legs. We
shall write
them in terms of non-degenerate $t$-distributions of lower orders and Feynman
propagators.

Some numerical distributions can be factorized [3,5] concerning their Lorentz
structure
$$t^{\nu 3\mu}_{uu\tilde u...}=g^{\nu\mu}\bar t^3_{uu\tilde u...},$$
$$t^{\nu 1\ldots}_{\psq\psi u\ldots}=\gamma^{\nu}\bar t^{1\ldots}_{\psq\psi
u\ldots},\>\>\>
t^{\nu 2\ldots}_{\psq\psi u\ldots}=\bar t^{2\ldots}_{\psq\psi
u\ldots}\gamma^{\nu},\eqno(1.22)$$
or concerning their colour structure
$$t_{B_1B_2\,ab}=\delta_{ab}t'_{B_1B_2},\>\>\>\>\>\>\>\>\>\>
t_{B_1B_2B_3\,abc}=f_{abc}t'_{B_1B_2B_3},$$
$$t_{\psq\psi\,\alpha\beta}=\tau_{\psq\psi}\delta_{\alpha\beta},\>\>\>\>
\>\>\>\>\>
t_{\psq\psi B\,a\,\alpha\beta}=\tau_{\psq\psi
B}(\lambda_a)_{\alpha\beta},\eqno(1.23)$$
where $B,B_1,B_2,B_3\in\{A,F,u,\tilde u\}$ and $a,b,c$
(rsp. $\alpha,\beta$) are indices of the adjoint (rsp. fundamental)
representation of SU(N).
We shall omit the prime in $t'_{B_1B_2(B_3)}$. Due to translation invariance,
the numerical
distributions depend on the relative coordinates
$(x_1-x_n,x_2-x_n,...,x_{n-1}-x_n)$ only.
The Fourier transformation is done in the latter coordinates and the hat in
$\hat t(p_1,p_2,...,p_{n-1})$ is mostly omitted.
\vskip 0.5cm
{\it (d) Assumption about the Infrared Behaviour}
\vskip 0.5cm
In the present paper we work with the following assumption throughout: Let
$p_1,...,p_r$
be the external and $p_{r+1},...,p_{n-1}$ be the inner momenta of an arbitrary
$\hat t$-distribution without or with one Q-vertex. Moreover
we set $p_n\=d -(p_1+p_2+...+p_{n-1})$. We assume that for
$$p_{r+1},...,p_{n-1},p_n\rightarrow 0,\>\>\>{\rm
and\>\>for\>\>all}\>\>p_1,...,p_{r-1}\>\>
{\rm and}\>\>\tilde p_r\=d -(p_1+...+p_{r-1})\>\>{\rm off-shell}\eqno(1.24)$$
the limit of the considered $\hat t$-distribution exists
$$\lim_{p_{r+1},...,p_{n-1},p_n\to 0}\hat t(p_1,p_2,...,p_{n-1})=
\hat t(p_1,...,p_r,0,...0)\vert_{p_1+...+p_r=0}.\eqno(1.25)$$
The mass-shell is meant to be the set $\{p\in {\bf R}^4|p^2=m^2\}$.
It is absolutely necessary that the external momenta are off-shell
because otherwise infrared divergences appear (even in QED [14,15]) and the
limit does not exist.

Remarks: (1) The vacuum diagrams ($r=0$) are excluded in (1.25). At the end of
the appendix it
is shown that they really violate this assumption.

(2) The limit (1.24-25) is understood in the following sense:
We start in $x$-space and smear out the
{\it inner} vertices with $g\in{\cal S}({\bf R}^4)$
$$\int d^4x_{r+1}...d^4x_n\>t(x_1-x_n,...,x_{n-1}-x_n)g(x_{r+1})...g(x_n)
\in{\cal S}'({\bf R}^{4r}).\eqno(1.26)$$
Next we consider the adiabatic limit $g\rightarrow 1$ of (1.26) in the sense of
tempered distributions
in $x_1,...,x_r$. For this purpose we replace $g(x)$ by
$$g_\epsilon (x)\=d g_0(\epsilon x),\>\>\>\>\>g_0(0)=1,\eqno(1.27)$$
$g_0\in{\cal S}({\bf R}^4)$ fixed, and consider $\epsilon\rightarrow
0,\,\epsilon >0$. Note
$$\hat g_\epsilon (k)=\epsilon^{-4}\hat g_0 ({k\over\epsilon}).\eqno(1.28)$$
Moreover we perform the Fourier transformation in the relative coordinates
$y_1\=d x_1-x_r,...,y_{r-1}\=d x_{r-1}-x_r$
$${1\over (2\pi)^{2(r-1)}}\int
d^4y_1...d^4y_{r-1}\>e^{i(p_1y_1+...+p_{r-1}y_{r-1})}
\int d^4x_{r+1}...d^4x_n$$
$$t(x_1-x_n,...,x_{n-1}-x_n)g_\epsilon(x_{r+1})...g_\epsilon(x_n)=$$
$$=\int d^4k_{r+1}...d^4k_n\>\hat g_0(k_{r+1})...\hat g_0(k_n)
e^{-i\epsilon(k_{r+1}+...+k_n)x_r}$$
$$\hat t(p_1,...,p_{r-1},-(p_1+...+p_{r-1})+\epsilon(k_{r+1}+...+k_n),-\epsilon
k_{r+1},
...,-\epsilon k_{n-1}),\eqno(1.29)$$
and obtain a tempered distribution in $p_1,...,p_{r-1},x_r$. The precise
formulation of
our assumtion (1.24-25) reads: Every point $p\=d (p_1,...,p_{r-1})$ with all
$p_1,...,p_{r-1}$
and with $\tilde p_r\=d -(p_1+...+p_{r-1})$ off-shell has a neighbourhood $U_p$
such that
in (1.29) the adiabatic limit $\epsilon\rightarrow 0$ exists
on the space of test functions with support in $(p,x_r)\in U_p\times {\bf
R}^4$.

(3) Our assumption (1.24-25) implies that the divergences with respect to inner
vertices
vanish in the infrared limit (1.24)
$$\lim_{p_{r+1},...,p_{n-1},p_n\to 0}p_{l\alpha}\hat t^{\alpha
l...}_{...}(p_1,...,p_{n-1})=0,
\quad\quad\forall l=r+1,...,n.\eqno(1.30)$$
We shall see in sect.2 that $p_{l\alpha}\hat t^{\alpha
l...}_{...}(p_1,...,p_{n-1}),\>
l=r+1,...,n$, which contains one Q-vertex, can {\it only} be eliminated from
the Cg-identities
by taking the limit (1.30).

(4) To prove the Cg-identities in [4] it was necessary to assume (1.30) and
that the infrared limit
(1.24-25) of certain (not all) $\hat t$-distributions rsp. combinations of
$\hat t$-distributions
exists. However, we approached this limit there from totally space-like points
only,
where the $\hat t$-distributions are analytic. Especially we merely considered
the region
$p_1^2,...,p_{r-1}^2,\tilde p_r^2\leq -\delta <0$. This was sufficient, since
we had to
investigate a polynomial only, namely the possible violation of a certain
Cg-identity.
We see that our assumption (1.24-25) is quite {\it stronger} than the one [4]
needed to prove the Cg-identities.

(5) In sect.3 we consider connected $r$-point {\it Green's functions} with the
external momenta
$k_1,k_2,...,k_r$, where $k_1+k_2+...+k_r=0$. Their perturbative expansions
will be written
in terms of our $\hat t$-distributions by means of the Gell-Mann Low series. In
doing so,
{\it all inner and the sum of the external momenta of the $\hat
t$-distributions are set equal
to zero}.
Considering the region with all $k_I\=d\sum_{i\in I}k_i$
off-shell (where $I$ runs through all subsets of $\{1,2,..,r\}$ with
$1,2,...,r-1$ elements),
the individual terms in these series exist if and only if our assumption
(1.24-25) holds true
for the corresponding $\hat t$-distributions. This will be evident by
considering the
explicit formulas. Usually the existence of the Gell-Mann Low expressions in
the mentioned region
is implicitly assumed in the literature. {\it A perturbative formulation
of the Slavnov-Taylor identities is impossible without our assumption
(1.24-25)}.
\vskip 1cm
{\trm 2. Elimination of the Distributions with one\break\vskip 0.3cm
Q-Vertex in the Cg-Identities}
\vskip 1cm
{\it (a) Two-Legs Cg-Identities}
\vskip 0.5cm
The 2-legs Cg-identities are derived and proven in sect.3 of [2].
Since numerical distributions
with an external $A$- or $F$-line at the same vertex get combined in Green's
functions
or matrix elements, we define
$$\tilde t^{\alpha\nu}_{AA}(p_1,p_2,...p_{n-1})\=d
t^{\alpha\nu}_{AA}(p_1,p_2,...)
+2ip_{1\tau} t^{\tau\alpha\nu}_{FA}(p_1,p_2,...)
+2ip_{2\tau} t^{\alpha\tau\nu}_{AF}(p_1,p_2,...)-$$
$$-4p_{1\tau}p_{2\rho} t^{\tau\alpha\rho\nu}_{FF}(p_1,p_2,...)\eqno(2.1)$$
and
$$\tilde t^{\alpha l\nu}_{uA}(p_1,p_2,...p_{n-1})\=d  t^{\alpha
l\nu}_{uA}(p_1,p_2,...)
+2ip_{2\tau} t^{\alpha l\tau\nu}_{uF}(p_1,p_2,...).\eqno(2.2)$$
In order to eliminate the distributions with one Q-vertex, we insert the 2-legs
Cg-identities
into each other and obtain
$$0=p_{1\alpha}\tilde t^{\alpha\nu}_{AA}(p_1,p_2,...)-i[p^2_2 t^\nu_{u\tilde u}
(p_1,p_2,...)-p_{2\mu}p_2^\nu  t^\mu_{u\tilde u}(p_1,p_2,...)]
+\sum_{l=3}^n p_{l\alpha}\tilde t^{\alpha l\nu}_{uA}(p_1,p_2,...),\eqno(2.3)$$
where $p_n\=d -(p_1+p_2+...+p_{n-1})$. We recognize that the $\tilde
t$-distributions
(2.1), (2.2) are well suited for this combined Cg-identity
(2.3). The latter equation still contains distributions with one Q-vertex,
namely in the
divergences with respect to inner vertices $\sum_{l=3}^n p_{l\alpha}\tilde
t^{\alpha l\nu}
_{uA}(p_1,p_2,...)$. In order to get rid of them we consider the infrared limit
(1.24):
$p_3\rightarrow 0,\,p_4\rightarrow 0,\,...\,p_n\rightarrow 0$, with
$p_1$ off-shell, and use (1.30). Taking additionally the covariant
decomposition
$$ t^\mu_{u\tilde u}(p,-p,0...)=p^\mu \bar t_{u\tilde u}(p^2)\eqno(2.4)$$
into account, we obtain for (2.3) the following simple reduced Cg-identity
$$0=p_{\alpha}\tilde t^{\alpha\nu}_{AA}(p,-p,0...).\eqno(2.5)$$
\vskip 0.5cm
{\it (b) Three-Legs Cg-Identities}
\vskip 0.5cm
We proceed in the same way for the 3-legs identities. First we consider the
case with
a pair $(\psq,\psi)$ among the external legs. The corresponding Cg-identities
are given and proven
in sect.4 of [5]. Defining analogously to (2.1)
$$\tilde\tau^\nu_{\psq\psi A}(p_1,p_2,p_3,p_4...,p_{n-1})\=d \tau^\nu_{\psq\psi
A}
(p_1,p_2,p_3,p_4...)+
2ip_{3\alpha}\tau^{\alpha\nu}_{\psq\psi F}(p_1,p_2,p_3,p_4...),\eqno(2.6)$$
we get by inserting (4.27) (of [5]) into (4.32) (of [5])
$$0=-ip_{3\nu}\tilde\tau^\nu_{\psq\psi A}(p_1,p_2,p_3,p_4...)-i\sum_{l=4}^n
p_{l\nu}
\tau^{\nu l}_{\psq\psi u}(p_1,p_2,p_3,p_4...)-$$
$$-i(\gamma^\nu p_{1\nu}-m)\bar\tau^{1}_{\psq\psi u}(p_1,p_2,p_3,p_4,...)
-i\bar\tau^{2}_{\psq\psi u}(p_1,p_2,p_3,p_4,...)(\gamma^\nu p_{2\nu}+m)+$$
$$+{g\over 2(2\pi)^2}[i\tau_{\psq\psi}(p_1+p_3,p_2,p_4,...)-i\tau_{\psq\psi}
(p_1,p_2+p_3,p_4,...)+\gamma_\mu t^\mu_{u\tilde
u}(p_3,p_1+p_2,p_4,...)],\eqno(2.7)$$
where again $p_n\=d -(p_1+p_2+...+p_{n-1})$. The divergences with respect to
inner
vertices, each containing one Q-vertex, can be eliminated by taking the
infrared
limit
$$p_4\rightarrow 0,\,...\,p_{n-1}\rightarrow 0,\,p_n\rightarrow 0,
\>\>\>\>\>\>\>p_1,p_2,p_3\>\>{\rm off-shell},\eqno(2.8)$$
and using (1.30)
$$0=-ip_{3\nu}\tilde\tau^\nu_{\psq\psi A}(p_1,p_2,p_3,0...)
-i(\gamma^\nu p_{1\nu}-m)\bar\tau^{1}_{\psq\psi u}(p_1,p_2,p_3,0,...)-$$
$$-i\bar\tau^{2}_{\psq\psi u}(p_1,p_2,p_3,0,...)(\gamma^\nu p_{2\nu}+m)+
{g\over 2(2\pi)^2}[i\tau_{\psq\psi}(p_1+p_3,p_2,0,...)-$$
$$-i\tau_{\psq\psi}
(p_1,p_2+p_3,0,...)+\gamma_\mu t^\mu_{u\tilde
u}(p_3,p_1+p_2,0,...)]\vert_{p_1+p_2+p_3=0}.
\eqno(2.9)$$
This reduced Cg-identity
still contains distributions with one Q-vertex: $\bar\tau^{1}_{\psq\psi u},
\,\bar\tau^{2}_{\psq\psi u}$. It is not possible to eliminate them, except the
external momenta $p_1,\,p_2$ approach the mass-shell: $\gamma^\nu
p_{1\nu}\rightarrow m,
\,\gamma^\nu p_{2\nu}\rightarrow -m$. But in the latter limit infrared
divergences appear.
Therefore, we avoid it.

We turn to the Cg-identities without external matter lines, derived in [3] and
proven in
[4]. We define
$$\tilde t^{\alpha\mu}_{Au\tilde u}(p_1,p_2,p_3,p_4...,p_{n-1})\=d
 t^{\alpha\mu}_{Au\tilde u}(p_1,p_2,p_3,p_4...)+2ip_{1\beta}
 t^{\beta\alpha\mu}_{Fu\tilde u}(p_1,p_2,p_3,p_4...),\eqno(2.10)$$
$$\tilde t^{\alpha\mu\nu}_{AAA}(p_1,p_2,p_3,p_4...,p_{n-1})\=d
t^{\alpha\mu\nu}_{AAA}(p_1,p_2,p_3,...)+2ip_{1\rho}
t^{\rho\alpha\mu\nu}_{FAA}(p_1,p_2,p_3,...)+$$
$$+2ip_{2\rho}t^{\alpha\rho\mu\nu}_{AFA}(p_1,p_2,p_3,...)
+2ip_{3\rho}t^{\alpha\mu\rho\nu}_{AAF}(p_1,p_2,p_3,...)
-4p_{1\rho}p_{2\tau}t^{\rho\alpha\tau\mu\nu}_{FFA}(p_1,p_2,p_3,...)-$$
$$-4p_{1\rho}p_{3\tau}t^{\rho\alpha\mu\tau\nu}_{FAF}(p_1,p_2,p_3,...)-
4p_{2\rho}p_{3\tau}t^{\alpha\rho\mu\tau\nu}_{AFF}(p_1,p_2,p_3,...)-$$
$$-8ip_{1\rho}p_{2\tau}p_{3\lambda}t^{\rho\alpha\tau\mu\lambda\nu}_{FFF}(p_1,
p_2,p_3,...)-$$
$$-{2g\over (2\pi)^2}\bigl[(t^{\alpha\nu\mu}_{AF}(p_1,p_2+p_3,p_4,...,p_{n-1})+
2ip_{1\tau}t^{\tau\alpha\nu\mu}_{FF}(p_1,p_2+p_3,p_4,...))+$$
$$+(t^{\mu\alpha\nu}_{AF}(p_2,p_1+p_3,p_4,...)+
2ip_{2\tau}t^{\tau\mu\alpha\nu}_{FF}(p_2,p_1+p_3,p_4,...))+$$
$$+(t^{\nu\mu\alpha}_{AF}(p_3,p_1+p_2,p_4,...)+
2ip_{3\tau}t^{\tau\nu\mu\alpha}_{FF}(p_3,p_1+p_2,p_4,...))\bigr]\eqno(2.11)$$
and
$$\tilde t^{\alpha l\mu\nu}_{uAA}(p_1,p_2,p_3,p_4...,p_{n-1})\=d
 t^{\alpha l\mu\nu}_{uAA}(p_1,p_2,p_3,p_4...)+
2ip_{2\rho}t^{\alpha l\rho\mu\nu}_{uFA}(p_1,p_2,p_3,p_4,...)+$$
$$+2ip_{3\rho}t^{\alpha l\mu\rho\nu}_{uAF}(p_1,p_2,p_3,p_4...)
-4p_{2\rho}p_{3\tau}t^{\alpha l\rho\mu\tau\nu}_{uFF}(p_1,p_2,p_3,p_4...)+$$
$$+{2g\over (2\pi)^2}t^{\alpha
(l-1)\mu\nu}_{uF}(p_1,p_2+p_3,p_4,...).\eqno(2.12)$$
Note that the 2-legs distributions in (2.11), (2.12) are of order $(n-1)$.
(2.10) and the 3-legs terms in (2.11), (2.12) are motivated by the combination
of
external $A$- and $F$-lines at the same vertex, similarly to (2.1), (2.2) and
(2.6).
The 2-legs terms in (2.11), (2.12) are  terms with an external four-gluon
vertex (1.18). The
diagram for the term $\sim gt^{\alpha\nu\mu}_{AF}(p_1,p_2+p_3,p_4,..,.p_{n-1})$
is given in fig.1 in $x$-space: The four-gluon vertex at $x_2=x_3$ on the
l.h.s. is generated
by the normalization term $-{1\over 2}\delta (x_2-x_3)$
of the propagator $\d\d D_F(x_2-x_3)$ on the r.h.s..
Now we eliminate the distributions with one Q-vertex in the list of
3-legs Cg-identities given in sect.3 of [3]. For this purpose the
identities of type I have already been inserted into the identities of type II
in this list.
We eliminate $t^2_{uAF}$ by inserting (3.14) into (3.13) (antisymmetrized in
$(x_2/p_2,\,\mu)\leftrightarrow (x_3/p_3,\,\nu)$). Inserting then the resulting
equation
into (3.12) we get rid of $t^3_{uAA}$. Moreover, $t^2_{uA}$ in (3.12b) is
eliminated by applying
the 2-legs Cg-identity (3.20) of [2]. We finally obtain
$$0=-ip_{1\alpha}\tilde t^{\alpha\mu\nu}_{AAA}(p_1,p_2,p_3,p_4...)
-i\sum_{l=4}^n p_{l\alpha}\tilde t^{\alpha l\mu\nu}_{uAA}(p_1,p_2,p_3,p_4...)
+p^2_3 \tilde t^{\mu\nu}_{Au\tilde u}(p_2,p_1,p_3,p_4,...)-$$
$$-p_3^\nu p_{3\alpha}\tilde t^{\mu\alpha}_{Au\tilde u}(p_2,p_1,p_3,p_4,...)
-p^2_2 \tilde t^{\nu\mu}_{Au\tilde u}(p_3,p_1,p_2,p_4,...)
+p_2^\mu p_{2\alpha}\tilde t^{\nu\alpha}_{Au\tilde u}(p_3,p_1,p_2,p_4,...)+$$
$$+{g\over (2\pi)^2}\bigl[\tilde t^{\mu\nu}_{AA}(p_1+p_2,p_3,p_4,...)
-\tilde t^{\nu\mu}_{AA}(p_1+p_3,p_2,p_4,...)+i[(2p_3^\mu +p_2^\mu)
t^\nu_{u\tilde u}(p_1,p_2+p_3,p_4,...)-$$
$$-g^{\mu\nu}p_{3\alpha}t^\alpha_{u\tilde u}(p_1,p_2+p_3,p_4,...)
-(2p_2^\nu +p_3^\nu)t^\mu_{u\tilde u}(p_1,p_2+p_3,p_4,...)+
g^{\mu\nu}p_{2\alpha}t^\alpha_{u\tilde
u}(p_1,p_2+p_3,p_4...)]\bigr].\eqno(2.13)$$
Without introducing the $\tilde t$-distributions, this combined Cg-identity
would be much more complicated. This is another motivation for the above
definitions.
Again we take the limit (2.8) and apply (1.30). Using additionally (2.4) we get
the reduced Cg-identity
$$0=-ip_{1\alpha}\tilde t^{\alpha\mu\nu}_{AAA}(p_1,p_2,p_3,0,...)
+p^2_3 \tilde t^{\mu\nu}_{Au\tilde u}(p_2,p_1,p_3,0,...)
-p_3^\nu p_{3\alpha}\tilde t^{\mu\alpha}_{Au\tilde u}(p_2,p_1,p_3,0,...)-$$
$$-p^2_2 \tilde t^{\nu\mu}_{Au\tilde u}(p_3,p_1,p_2,0,...)
+p_2^\mu p_{2\alpha}\tilde t^{\nu\alpha}_{Au\tilde u}(p_3,p_1,p_2,0,...)+$$
$$+{g\over (2\pi)^2}\bigl[\tilde t^{\mu\nu}_{AA}(-p_3,p_3,0,...)
-\tilde t^{\nu\mu}_{AA}(-p_2,p_2,0,...)+$$
$$+i\bar t_{u\tilde u}(p^2_1)[p^2_3 g^{\mu\nu}-p^\mu_3 p^\nu_3
-p^2_2 g^{\mu\nu}+p^\mu_2 p^\nu_2]\bigr]\vert_{p_1+p_2+p_3=0}.\eqno(2.14)$$
Only distributions of the physical theory (i.e. distributions without Q-vertex)
appear in this equation.

There remains the 3-legs Cg-identity (3.9) of [3].
Going again over to the infrared limit (2.8), this Cg-identity is reduced by
means of (1.30) to
$$0=-ip_{1\alpha}\tilde t^{\alpha\mu}_{Au\tilde u}(p_1,p_2,p_3,0...)
-ip_{2\alpha}\tilde t^{\alpha\mu}_{Au\tilde u}(p_2,p_1,p_3,0...)
-ip^\mu_3\bar t^3_{uu\tilde u}(p_1,p_2,p_3,0...)+$$
$$+{g\over (2\pi)^2}[t_{u\tilde u}^\mu (p_1+p_2,p_3,0...)
-t_{u\tilde u}^\mu (p_2,p_1+p_3,0...)-t_{u\tilde u}^\mu (p_1,p_2+p_3,0...)]
\vert_{p_1+p_2+p_3=0}.\eqno(2.15)$$
However, $\bar t^3_{uu\tilde u}$, which contains one Q-vertex, cannot be
eliminated.
\vskip 0.5cm
{\it (c) Four-Legs Cg-Identities}
\vskip 0.5cm
To shorten the story we immediately take the limit
$$p_5,\,...\,p_{n-1},\,p_n\rightarrow 0,
\>\>\>\>\>\>\>p_k,\>(p_i+p_j)\>\>{\rm off-shell}\>\>
\forall k=1,...,4,\>\>\forall 1\leq i<j\leq 4.\eqno(2.16)$$
Besides $p_k$, the momenta $(p_i+p_j)$ need to be off-shell too, since they
appear
in the arguments of the 3- or 2-legs distributions in the degenerate terms.
(The latter
include terms with external four-gluon vertex/vertices.)
Due to (1.30), the divergences
with respect to inner vertices vanish in this limit (2.16). We define
$$\tilde t^{\nu\mu}_{\psq\psi AA\,ab}(p_1,p_2,p_3,p_4,p_5,...,p_{n-1})\=d
t^{\nu\mu}_{\psq\psi AA\,ab}(p_1,...)+2ip_{3\alpha}t^{\alpha\nu\mu}_{\psq\psi
FA\,ab}(p_1,...)+$$
$$+2ip_{4\alpha}t^{\nu\alpha\mu}_{\psq\psi
AF\,ab}(p_1,...)-4p_{3\alpha}p_{4\beta}
t^{\alpha\nu\beta\mu}_{\psq\psi FF\,ab}(p_1,...)+{2g\over (2\pi)^2}
\tau^{\nu\mu}_{\psq\psi
F}(p_1,p_2,p_3+p_4,p_5,...)f_{abc}\lambda_c,\eqno(2.17)$$
$$\tilde t^{1\mu}_{\psq\psi uA}(p_1,p_2,p_3,p_4,p_5,...,p_{n-1})\=d
\bar t^{1\mu}_{\psq\psi uA}(p_1,...)+2ip_{4\alpha}\bar t^{1\alpha\mu}_{\psq\psi
uF}(p_1,...),
\eqno(2.18)$$
$$\tilde t^{2\mu}_{\psq\psi uA}(p_1,p_2,p_3,p_4,p_5,...,p_{n-1})\=d
\bar t^{2\mu}_{\psq\psi uA}(p_1,...)+2ip_{4\alpha}\bar t^{2\alpha\mu}_{\psq\psi
uF}(p_1,...),
\eqno(2.19)$$
$$\tilde t^{\tau\nu\mu}_{u\tilde u
AA\,abcd}(p_1,p_2,p_3,p_4,p_5,...,p_{n-1})\=d
t^{\tau\nu\mu}_{u\tilde u AA\,abcd}(p_1,...)
+2ip_{3\alpha}t^{\tau\alpha\nu\mu}_{u\tilde uFA\,abcd}(p_1,...)+$$
$$+2ip_{4\alpha}t^{\tau\nu\alpha\mu}_{u\tilde u
AF\,abcd}(p_1,...)-4p_{3\alpha}p_{4\beta}
t^{\tau\alpha\nu\beta\mu}_{u\tilde u FF\,abcd}(p_1,...)+{2g\over (2\pi)^2}
t^{\tau\nu\mu}_{u\tilde
uF}(p_1,p_2,p_3+p_4,p_5,...)f_{abr}f_{cdr},\eqno(2.20)$$
$$\tilde
t^{\alpha\nu\kappa\lambda}_{AAAA\,abcd}(p_1,p_2,p_3,p_4,p_5,...,p_{n-1})\=d
t^{\alpha\nu\kappa\lambda}_{AAAA\,abcd}(p_1,...)
+2ip_{1\beta}t^{\beta\alpha\nu\kappa\lambda}_{FAAA\,abcd}(p_1,...)+$$
$$+2ip_{2\gamma}t^{\alpha\gamma\nu\kappa\lambda}_{AFAA\,abcd}(p_1,...)
+2ip_{3\rho}t^{\alpha\nu\rho\kappa\lambda}_{AAFA\,abcd}(p_1,...)
+2ip_{4\tau}t^{\alpha\nu\kappa\tau\lambda}_{AAAF\,abcd}(p_1,...)-$$
$$-4p_{1\beta}p_{2\gamma}t^{\beta\alpha\gamma\nu\kappa\lambda}_{FFAA\,abcd}
(p_1,...)
-4p_{1\beta}p_{3\rho}t^{\beta\alpha\nu\rho\kappa\lambda}_{FAFA\,abcd}
(p_1,...)
-4p_{1\beta}p_{4\tau}t^{\beta\alpha\nu\kappa\tau\lambda}_{FAAF\,abcd}
(p_1,...)-$$
$$-4p_{2\gamma}p_{3\rho}t^{\alpha\gamma\nu\rho\kappa\lambda}_{AFFA\,abcd}
(p_1,...)
-4p_{2\gamma}p_{4\tau}t^{\alpha\gamma\nu\kappa\tau\lambda}_{AFAF\,abcd}
(p_1,...)-$$
$$-4p_{3\rho}p_{4\tau}t^{\alpha\nu\rho\kappa\tau\lambda}_{AAFF\,abcd}(p_1,...)
-8ip_{1\beta}p_{2\gamma}p_{3\rho}t^{\beta\alpha\gamma\nu\rho\kappa\lambda}_
{FFFA\,abcd}(p_1,...)-$$
$$-8ip_{1\beta}p_{2\gamma}p_{4\tau}t^{\beta\alpha\gamma\nu\kappa\tau\lambda}_
{FFAF\,abcd}(p_1,...)
-8ip_{1\beta}p_{3\rho}p_{4\tau}t^{\beta\alpha\nu\rho\kappa\tau\lambda}_{FAFF\
,abcd}(p_1,...)-$$
$$-8ip_{2\gamma}p_{3\rho}p_{4\tau}t^{\alpha\gamma\nu\rho\kappa\tau\lambda}_
{AFFF\,abcd}(p_1,...)
+16p_{1\beta}p_{2\gamma}p_{3\rho}p_{4\tau}t^{\beta\alpha\gamma\nu\rho\kappa
\tau\lambda}
_{FFFF\,abcd}(p_1,...)+$$
$$+{2g\over (2\pi)^2}\Bigl\{f_{abr}f_{cdr}[t^{\alpha\nu\kappa\lambda}_{AAF}
(p_1,p_2,p_3+p_4,p_5,...)+2ip_{1\beta}t^{\beta\alpha\nu\kappa\lambda}_{FAF}
(p_1,p_2,p_3+p_4,p_5,...)+$$
$$+2ip_{2\gamma}t^{\alpha\gamma\nu\kappa\lambda}_{AFF}(p_1,p_2,p_3+p_4,
p_5,...)
-4p_{1\beta}p_{2\gamma}t^{\beta\alpha\gamma\nu\kappa\lambda}_{FFF}(p_1,p_2,
p_3+p_4,p_5,...)+$$
$$+t^{\kappa\lambda\alpha\nu}_{AAF}(p_3,p_4,p_1+p_2,p_5,...)
+2ip_{3\rho}t^{\rho\kappa\lambda\alpha\nu}_{FAF}(p_3,p_4,p_1+p_2,p_5,...)+$$
$$+2ip_{4\tau}t^{\kappa\tau\lambda\alpha\nu}_{AFF}(p_3,p_4,p_1+p_2,p_5,...)
-4p_{3\rho}p_{4\tau}t^{\rho\kappa\tau\lambda\alpha\nu}_{FFF}(p_3,p_4,p_1+
p_2,p_5,...)+$$
$$+{2g\over
(2\pi)^2}t^{\alpha\nu\kappa\lambda}_{FF}(p_1+p_2,p_3+p_4,p_5,...)]\Bigl\}+$$
$$+{2g\over (2\pi)^2}\Bigl\{{\rm one\> cyclic\> permutation}\>
(b,p_2,\nu)\rightarrow
(c,p_3,\kappa)\rightarrow (d,p_4,\lambda)\rightarrow $$
$$\rightarrow (b,p_2,\nu)\Bigl\}
+{2g\over (2\pi)^2}\Bigl\{{\rm two\> cyclic\> permutations}\Bigl\}\eqno(2.21)$$
and
$$\tilde t^{3\nu}_{uu\tilde uA}(p_1,...,p_{n-1})\=d
\bar t^{3\nu}_{uu\tilde uA}(p_1,...,p_{n-1})
+2ip_{4\tau}\bar t^{3\tau\nu}_{uu\tilde uF}(p_1,...,p_{n-1}).\eqno(2.22)$$
The motivation for these definitions is the same as above. The terms with a
3-legs
distribution are terms with one external four-gluon vertex, analogously to
fig.1.
The $t_{FF}$-terms in $\tilde t_{AAAA}$ have two external four-gluon vertices:
The diagram in x-spaxe belonging to the term $\sim
g^2t_{FF}(p_1+p_2,p_3+p_4,p_5,...)$
has one four-gluon vertex at $x_1=x_2$ and another at $x_3=x_4$.

First we consider the 4-legs Cg-identities with one pair $(\psq,\psi)$ among
the external legs.
They are derived and proven in sect.4 of [5]. Inserting the identities of type
I
into the identities of type II and (4.34) (of [5]) into (4.33) (of [5]) we
obtain
the reduced Cg-identity
$$0=-ip_{3\nu}\tilde t^{\nu\mu}_{\psq\psi AA\,ab}(p_1,p_2,p_3,p_4,0,...)-$$
$$-i(\gamma^\nu p_{1\nu}-m)\tilde t^{1\mu}_{\psq\psi
uA\,ab}(p_1,p_2,p_3,p_4,0,...)
-i\tilde t^{2\mu}_{\psq\psi uA\,ab}(p_1,p_2,p_3,p_4,0,...)(\gamma^\nu
p_{2\nu}+m)+$$
$$+p^\mu_4p_{4\nu}t^{\nu}_{\psq\psi u\tilde u\,ab}(p_1,p_2,p_3,p_4,0,...)
-p^2_4t^{\mu}_{\psq\psi u\tilde u\,ab}(p_1,p_2,p_3,p_4,0,...)+$$
$$+{g\over (2\pi)^2}\Bigl\{{i\over
2}[\lambda_a\lambda_b\tilde\tau^\mu_{\psq\psi A}
(p_1+p_3,p_2,p_4,0,...)-\lambda_b\lambda_a\tilde\tau^\mu_{\psq\psi
A}(p_1,p_2+p_3,p_4,0,...)]-$$
$$-{1\over 2}\gamma_\lambda f_{abc}\lambda_c\tilde t^{\mu\lambda}_{Au\tilde
u}(p_4,p_3,p_1+p_2,0,...)
+f_{abc}\lambda_c\tilde\tau^\mu_{\psq\psi A}(p_1,p_2,p_3+p_4,0,...)-$$
$$-{i\over 2}[\gamma^\mu\lambda_b\lambda_a\bar\tau^1_{\psq\psi
u}(p_1+p_4,p_2,p_3,0,...)
-\bar\tau^2_{\psq\psi
u}(p_1,p_2+p_4,p_3,0,...)\lambda_a\lambda_b\gamma^\mu]\Bigl\}
\vert_{p_1+p_2+p_3+p_4=0}.\eqno(2.23)$$
The distributions $\tilde t^{1\mu}_{\psq\psi uA},\,\tilde t^{2\mu}_{\psq\psi
uA},\,
\bar\tau^1_{\psq\psi u},\,\bar\tau^2_{\psq\psi u}$, which have one Q-vertex,
cannot be eliminated.

The 4-legs Cg-identities without external $\psq,\psi$ are given in sect.4 of
[3] and
proven in [4]. We eliminate $t^2_{uAFF}$ by inserting (4.28) (of [3]) into
(4.27). Then
we insert the equation obtained in this way into (4.24), to get rid of
$t^3_{uAAF}$.
By substituting the resulting equation into (4.17), we eliminate $t^2_{uAAA}$.
The 3-legs terms
still contain distributions with one Q-vertex. By using (3.13) of [3] we
eliminate
$t^2_{uAA}$ and then we get rid of $t^2_{uAF}$ by means of (3.14) (of [3]). The
2-legs terms
$\sim t_{AF}$ and $\sim t_{FF}$ cancel due to Jacobi's identity. The result of
this lengthy
insertions is a reduced Cg-identity without any distribution with Q-vertex
$$0=-ip_{1\alpha}\tilde
t^{\alpha\nu\kappa\lambda}_{AAAA\,abcd}(p_1,p_2,p_3,p_4,0,...)+$$
$$+\biggl[p_{2\alpha}p_2^\nu \tilde t^{\alpha\kappa\lambda}_{u\tilde u
AA\,abcd}
(p_1,p_2,p_3,p_4,0,...)-p^2_2 \tilde t^{\nu\kappa\lambda}_{u\tilde u AA\,abcd}
(p_1,p_2,p_3,p_4,0,...)+$$
$$+{g\over (2\pi)^2}f_{abr}f_{cdr}\Bigl\{\tilde t^{\nu\kappa\lambda}_{AAA}
(p_1+p_2,p_3,p_4,0,...)+i(2p_3+p_4)^\lambda \tilde t^{\nu\kappa}_{Au\tilde u}
(p_2,p_1,p_3+p_4,0,...)-$$
$$-i(2p_4+p_3)^\kappa \tilde t^{\nu\lambda}_{Au\tilde u}(p_2,p_1,p_3+p_4,0,...)
+ig^{\lambda\kappa}(p_4-p_3)_\alpha\tilde t^{\nu\alpha}_{Au\tilde
u}(p_2,p_1,p_3+p_4,0,...)+$$
$$+{g\over (2\pi)^2}[g^{\nu\kappa}t^\lambda_{u\tilde u}(p_1,-p_1,0,...)
-(\kappa \leftrightarrow \lambda)]\Bigr\}\biggr]+$$
$$+\biggl[{\rm one\> cyclic\> permutation}\> (b,p_2,\nu)\rightarrow
(c,p_3,\kappa)\rightarrow (d,p_4,\lambda)\rightarrow $$
$$\rightarrow (b,p_2,\nu)\biggr]
+\biggl[{\rm two\> cyclic\>
permutations}\biggr]\vert_{p_1+p_2+p_3+p_4=0}.\eqno(2.24)$$

There remain the non-trivial 4-legs Cg-identities (4.20) and (4.29) in sect.4
of [3].
By inserting the latter into the first one, we eliminate $t^4_{uu\tilde uA}$.
Moreover we eliminate $\bar t^3_{uu\tilde u}$ by means of (2.15)
$$0=-i[p_{1\alpha}\tilde t^{\mu\alpha\nu}_{u\tilde
uAA\,bcad}(p_2,p_3,p_1,p_4,0,...)]
+i[(a,p_1)\leftrightarrow (b,p_2)]-$$
$$-ip_3^\mu \tilde t^{3\nu}_{uu\tilde uA\,abcd}(p_1,p_2,p_3,p_4,0,...)+$$
$$+p^\nu_4p_{4\beta}t^{\mu\beta}_{uu\tilde u\tilde
u\,abcd}(p_1,p_2,p_3,p_4,0,...)
-p^2_4t^{\mu\nu}_{uu\tilde u\tilde u\,abcd}(p_1,p_2,p_3,p_4,0,...)+$$
$$+{g\over (2\pi)^2}\biggl\{f_{abr}f_{cdr}\bigl [\tilde t^{\nu\mu}_{Au\tilde u}
(p_4,p_1+p_2,p_3,0,...)+$$
$$+{g^{\nu\mu}\over (p_1+p_2)^2}\{p_{1\alpha}(p_1+p_2)_\beta
\tilde t^{\alpha\beta}_{Au\tilde u}(p_1,p_2,p_3+p_4,0,...)+
p_{2\alpha}(p_1+p_2)_\beta\tilde t^{\alpha\beta}_{Au\tilde
u}(p_2,p_1,p_3+p_4,0,...)+$$
$$+{ig\over (2\pi)^2}[(p_1+p_2)^2\bar t_{u\tilde u}((p_1+p_2)^2)-(p_1p_2+p_2^2)
\bar t_{u\tilde u}(p_2^2)-(p_1p_2+p_1^2)\bar t_{u\tilde u}(p_1^2)]\}\bigr ]+$$
$$+f_{adr}f_{bcr}[\tilde t^{\nu\mu}_{Au\tilde u}(p_1+p_4,p_2,p_3,0,...)
+\tilde t^{\nu\mu}_{Au\tilde u}(p_4,p_1,p_2+p_3,0,...)]+$$
$$+f_{acr}f_{dbr}[\tilde t^{\nu\mu}_{Au\tilde u}(p_2+p_4,p_1,p_3,0,...)
+\tilde t^{\nu\mu}_{Au\tilde u}(p_4,p_2,p_1+p_3,0,...)]\biggr\}
\vert_{p_1+p_2+p_3+p_4=0}.\eqno(2.25)$$
This equation still contains a distribution with one Q-vertex, namely
$\tilde t^{3}_{uu\tilde uA}$, which cannot be eliminated.
Again, the introduction of the $\tilde t$-distributions
simplifies the reduced Cg-identities (2.23-25) enormously.
\vskip 0.5cm
{\it (d) Gauge Invariant (Finite) Renormalizations}
\vskip 0.5cm
We study finite renormalizations in the above limits of vanishing inner momenta
(including
$p_n\=d -(p_1+...+p_{n-1})$) and with (sums of) the external momenta
off-shell.
Note that for the Gell-Mann Low series of the connected Green's functions
(sect.3)
we shall need the values of the $t$-distributions in these limits only.
The freedom of normalization (1.16) of a numerical distribution $\hat t$ is a
Lorentz covariant, SU(N)-invariant polynomial (in $p_1,p_2,...,p_{n-1}$) of
degree
$\omega$ which has the same permutation symmetries as $t$ [2,3,4,5],
where $\omega$ is given by (1.17). Inserting these
renormalization polynomials of the $t$'s into the definition of $\tilde t$
(e.g. into (2.1)),
we obtain the (undetermined) renormalization polynomial of
$\tilde t$. These are the results:
$$\tilde t^{\mu\nu}_{AA}(p,-p,0,...)+C^1_{AA}g^{\mu\nu}+C^2_{AA}g^{\mu\nu}p^2
+C^3_{AA}p^\mu p^\nu,\eqno(2.26)$$
$$t_{u\tilde u}^\nu (p,-p,0,...)+iC_{u\tilde u}p^\nu,\eqno(2.27)$$
$$\tau_{\psq\psi}(p,-p,0,...)+C^0_{\psq\psi}+C_{\psq\psi}\gamma^\nu
p_\nu,\eqno(2.28)$$
$$\tilde t^{\nu\kappa\lambda}_{AAA}(p_1,p_2,p_3,0,...)
-iC_{AAA}[g^{\nu\kappa} (p_1-p_2)^\lambda +g^{\nu\lambda}(p_3-p_1)^\kappa
+g^{\kappa\lambda}(p_2-p_3)^\nu]\vert_{p_1+p_2+p_3=0},\eqno(2.29)$$
$$\tilde t^{\mu\nu}_{Au\tilde u}+g^{\mu\nu}C_{Au\tilde u},\eqno(2.30)$$
$$\tilde\tau^\nu_{\psq\psi A}+C_{\psq\psi A}\gamma^\nu,\eqno(2.31)$$
$$\bar\tau^1_{\psq\psi u}+C_{\psq\psi u},\>\>\>\>\>\>\>\>
\bar\tau^2_{\psq\psi u}+C_{\psq\psi u},\eqno(2.32)$$
$$\bar t^3_{uu\tilde u}+C_{uu\tilde u},\eqno(2.33)$$
$$\tilde
t^{\alpha\nu\kappa\lambda}_{AAAA\,abcd}+C_{AAAA}\biggl\{[f_{abr}f_{cdr}
(-g^{\alpha\lambda}g^{\nu\kappa}+g^{\alpha\kappa}g^{\nu\lambda})]+$$
$$+[{\rm one\> cyclic\> permutation}\> (b,\nu)\rightarrow
(c,\kappa)\rightarrow (d,\lambda)\rightarrow (b,\nu)]+$$
$$+[{\rm two\> cyclic\>
permutations}]\biggr\}+N^{\alpha\nu\kappa\lambda}_{AAAA\,abcd}.\eqno(2.34)$$
The other 4-legs distributions cannot be renormalized, because their singular
order is
$\omega\leq -1$. The equality of the renormalization terms of
$\bar\tau^1_{\psq\psi u}$
and $\bar\tau^2_{\psq\psi u}$ in (2.32) is a consequence of C-invariance [5].
The colour space
of the 4-legs distributions was studied in [4] (sect.3(a) and appendix A). It
is 5-
or 6-dimensional. The linear span of the tensors $f_{..r}f_{..r}$ is a
2-dimensional
subspace. $N_{AAAA\,abcd}$ in (2.34) is an element of a linear complement of
this subspace.
For our purposes it is not necessary to determind the form
of $N_{AAAA\,abcd}$ explicitly.

We assume that the initial $t,\tilde t$-distributions fulfil our reduced
Cg-identities
and require that the renormalized ones fulfil them, too. This gives the
following restrictions
on the renormalization constants in (2.26-34): From (2.5)
$$C^1_{AA}=0,\>\>\>\>\>\>\>C^2_{AA}=-C^3_{AA}(\=d C_{AA}),\eqno(2.35)$$
from (2.9)
$$0=-C_{\psq\psi A}+C_{\psq\psi u}+{g\over 2(2\pi)^2}(C_{\psq\psi}+C_{u\tilde
u}),\eqno(2.36)$$
from (2.14)
$$0=-C_{AAA}+C_{Au\tilde u}+{g\over (2\pi)^2}(C_{AA}-C_{u\tilde
u}),\eqno(2.37)$$
from (2.15)
$$C_{uu\tilde u}=C_{Au\tilde u},\eqno(2.38)$$
from (2.23)
$$C_{\psq\psi u}=-{1\over 2}C_{Au\tilde u},\eqno(2.39)$$
from (2.24)
$$N^{\alpha\nu\kappa\lambda}_{AAAA\,abcd}=0,\>\>\>\>\>C_{AAAA}={g\over
(2\pi)^2}
(C_{AAA}+C_{Au\tilde u}-{g\over (2\pi)^2}C_{u\tilde u})\eqno(2.40)$$
and (2.25) implies no restriction, due to the Jacobi identity.
Assuming $C_{AAAA}\sim g^n$, the 3-legs (2-legs) normalization
constants are of order $(n-1)$ (rsp. $(n-2)$).

Every reduced Cg-identity (2.9), (2.23), (2.15) and (2.25) still contains one
Q-vertex.
However, there is a big difference between the first two and the latter two:
(2.15) and (2.25)
yield (2.38) only, which is physically of no importance, since $\bar
t^3_{uu\tilde u}$ (2.33)
has one Q-vertex. These two reduced Cg-identities are irrelevant for the
normalization of the physical
$t$-distributions. On the other hand, $C_{\psq\psi u}$, which belongs to the
unphysical
distributions $\bar\tau^1_{\psq\psi u},\,\bar\tau^2_{\psq\psi u}$ in (2.9),
(2.23),
can be eliminated by inserting (2.39) into (2.36). Therefore, (2.9) and (2.23)
give restrictions on the physical theory. Apparantly this was the motivation
for Taylor [10]
to study the corresponding Slavnov-Taylor identities (sect.3), although he had
to introduce the unphysical Q-vertex for this purpose. The result of the above
mentioned
insertion and (2.40) can be written in the form
$$0=-C_{\psq\psi A}-{1\over 2}C_{AAA}+{g\over
2(2\pi)^2}(C_{\psq\psi}+C_{AA}),\eqno(2.41)$$
$$C_{AAAA}={g\over (2\pi)^2}(2C_{AAA}-{g\over (2\pi)^2}C_{AA}),\eqno(2.42)$$
where (2.37) has been used.

Remarks: (A)  The renormalizations (2.26-34) with the restrictions (2.35-42)
are the most
general finite renormalizations of the $\tilde t$-distributions (rsp.
$t$-distributions
in the cases where no $\tilde t$ was defined) which conserve the {\it reduced}
Cg-identities. However, the latter contain {\it less} information than the
original Cg-identities [2,3,5]. Therefore, it could be possible that the
original
Cg-identities give {\it stronger} restrictions on the renormalizations
(2.26-34)
than (2.35-42). Fortunately this holds not true. Starting with an arbitrary
renormalization
(2.26-34) of the $\tilde t$- (rsp. $t$-) distributions fulfilling (2.35-42),
one can
find a corresponding renormalization of the $t$-distributions which preserves
the original Cg-identities. This can be done in the following way: Let
$N(p_1,...,p_r)\vert_{p_1+...+p_r=0}$ be a renormalization polynomial of
$\tilde t_{B_1...B_r}(p_1,...,p_r,0,...,0)\vert_{p_1+...+p_r=0}$. This
renormalization can
be generated by the renormalization
$t_{B_1...B_r}(p_1,...,p_{n-1})+N(p_1,...,p_r)$
and by keeping the other $t$-distributions (appearing in the definition of
$\tilde t_{B_1...B_r}$) unchanged. Together with a suitable renormalization of
the
$t^l_{uB_2...B_r}$-distributions, $l=r+1,...,n$, the original Cg-identities can
be maintained.

(B) By means of $Z_i\=d 1+\alpha_i C_i$ (where $\alpha_i$ is a suitable number,
e.g.
${g\over (2\pi)^2}$) one can define {\it finite} Z-factors. Omitting terms
quadratic in $C_i$,
the equations (2.35-42) become the well-known [10,11,16] Z-factor relations.
This is correct
at one-loop level only. For the definition of the Z-factors in higher orders,
the $C_i$
need to be normalization constants of one-particle irreducible diagrams.
However, our
$t,\tilde t$-distributions contain one-particle reducible terms. Therefore, and
because
the {\it multiplicative} renormalization of the Z-factors contrasts with our
{\it additive}
one (1.16), the Z-factors are not natural in our framework.
It is well-known [10,11,17] that the Slavnov-Taylor identities,
which we are going to prove by means of our reduced Cg-identities in the
following section,
imply the Z-factor relations in all orders.
\vskip 1cm
{\trm 3. Derivation of the Slavnov-Taylor Identities\break\vskip 0.3cm
from the reduced Cg-Identities}
\vskip 1cm
The Slavnov-Taylor identities are given in terms of {\it Green's functions}
(see some
conventional textbook)
$$<\Omega\mid T(\Phi_{j_1\,{\rm int}}(x_1)...\Phi_{j_r\,{\rm int}}(x_r))\mid
\Omega >,\eqno(3.1)$$
where $\Phi_{j_1\,{\rm int}}(x_1),...\Phi_{j_r\,{\rm int}}(x_r)$ are
interacting fields, $T$ means
the time ordering and $\mid \Omega >$ is the vacuum or ground state of the
interacting
theory. By means of the Gell-Mann Low series [13]
we know the perturbative expansion of (3.1) (see the examples below): Let
$k_1,k_2,...,k_r$ be the external momenta, where $k_1+k_2+...+k_r=0$. In order
to avoid
infrared divergences, we always assume all $k_I\=d\sum_{i\in I}k_i$ to be
off-shell
(where $I$ runs through all subsets of $\{1,2,..,r\}$ with $1,2,...,r-1$
elements).
Then, as we will see, our assumption (1.24-25) implies the existence of the
Gell-Mann Low expressions in the following sense: Every point $k\=d
(k_1,...,k_{r-1})$
in the above region has a neighbourhood $U_k\subset {\bf R}^{4(r-1)}$, such
that the
Gell-Mann Low expressions are distributions on the space of testfunctions with
support in $U_k$.
Because of the degenerate terms it does not suffice
to keep only $k_1,k_2,...,k_r$ off-shell.
For our purposes we only need the {\it connected} Green's functions
$<\Omega\mid T(\Phi_{j_1\,{\rm int}}(x_1)...\Phi_{j_r\,{\rm int}}(x_r))\mid
\Omega >_c$.
Their perturbative expansion is obtained by omitting the disconnected diagrams
in the
Gell-Mann Low series. This omission is no problem for us, since our $t,\,\tilde
t,
\,\tau$-distributions contain connected diagrams only.

Epstein and Glaser [6] give another definition of Green's functions in the
framework
of causal perturbation theory. In these expressions the inner vertices are
smeared
out with $g\in{\cal S}({\bf R}^4)$. Due to Epstein and Glaser [6] the adiabatic
limit $g\rightarrow 1$ exists for {\it massive} theories. In this
limit their Green's functions agree with the Gell-Mann Low expressions, if the
latter are infrared-improved in an appropriate way.
This is proven in the appendix for the massive case.
In our Yang-Mills model with matter fields the infrared behaviour is worse
[23]. Therefore, we
only consider the region with all $k_I$ (defined above)
off-shell. In this region the two definitions of Green's
functions probably agree if the adiabatic limit
is taken in the Epstein-Glaser expressions. In fact, in the appendix it is
demonstrated
by an explicit calculation that this is true at least for the 2-point Green's
functions
in lowest non-trivial order. Since the aim of this paper is to derive the {\it
usual}
Slavnov-Taylor identities from our Cg-identities, we work with the conventional
perturbative expansion of Green's functions - the (renormalized) Gell-Mann Low
series for (3.1).
\vskip 0.5cm
{\it (a) Two-Legs Slavnov-Taylor Identity}
\vskip 0.5cm
For the sake of completeness we summarize and discuss sect.3.3 of [2]. One
defines the 2-point Green's function or gluon propagator [16] by
$$iD_{ab}^{\mu\nu}(k)\equiv i\delta_{ab}D^{\mu\nu}(k)\=d (2\pi)^{-2}\int
d^4x\,e^{ikx}
<\Omega \mid T(A_{{\rm int}\,a}^\mu (x)A_{{\rm int}\,b}^\nu (0))\mid
\Omega>_c,\eqno (3.2)$$
where $A_{{\rm int}}$ is the interacting gluon field. (3.2) has the
following perturbative expansion
$$iD_{ab}^{\mu\nu}(k)=-i(2\pi)^{-2}\delta_{ab}{g^{\mu\nu}\over k^2+i0}-
\delta_{ab}{1\over k^2+i0}\sum_{n=1}^\infty {(2\pi )^{(4n-4)}\over (2n-2)!}
\tilde t_{AA}^{(2n)\,\mu\nu}(k,-k,0,...,0){1\over k^2+i0},\eqno (3.3)$$
where $\tilde t_{AA}^{(2n)}$ is defined in (2.1). Besides the limit (1.25) of
the
$\hat t$-distributions, there is a second source of infrared divergences in the
Gell-Mann Low series, namely the square of a Feynman propagator, which appears
in
the two-legs terms. Without an infrared regularization (e.g. the adiabatic
switching
with $g\in{\cal S}({\bf R}^4)$, see the appendix), the terms $n\geq 1$ in (3.3)
do not exist
(in the sense of distributions in $k$) in a neighbourhood of $k^2=0$, except
the normalization constants $C_{AA}^{(2n)}$
(2.35) of $\tilde t_{AA}^{(2n)}$ can be chosen in such a way that
$$\tilde t_{AA}^{(2n)\,\mu\nu}(k,-k,0,...,0)\sim k^2(k^\mu k^\nu
-g^{\mu\nu}k^2)
,\>\>\>\>\forall n\eqno(3.3a)$$
for $k^2\rightarrow 0$.
Such a normalization is given in QED by the central solution of the
distribution
splitting [6,8,18] and it exists also for the matter loop in $\tilde
t_{AA}^{(2)}$
(second order) if the matter fields are massive. However, the sum of the gluon
and ghost loop is [2]
$$\sim (k^\mu k^\nu -g^{\mu\nu}k^2) {\rm log}{-(k^2+i0)\over M^2}\eqno(3.3b)$$
in second order, where $M$ is an arbitrary constant. (The choice of $M$ is the
choice of
$C_{AA}^{(2)}$ (2.35).) Obviously (3.3a) is impossible. We avoid serious
problems by assuming
throughout all (sums of) external momenta to be off-shell.

By means of (2.5) we obtain the well-known Slavnov-Taylor identity [11,16]
$$k_\mu D_{ab}^{\mu\nu}(k)=-(2\pi)^{-2}\delta_{ab}{k^\nu\over k^2}\eqno(3.4)$$
and this is the {\it only} one with two legs.
\vskip 0.5cm
{\it (b) Three-Legs Slavnov-Taylor Identities}
\vskip 0.5cm
Analogous to (3.1) one defines [16] the ghost propagator by
$$i\tilde D_{ab}(k)\equiv i\delta_{ab}\tilde D(k):\=d (2\pi)^{-2}
\int d^4x\,e^{ikx}<\Omega \mid T(u_{{\rm int}\,a}
(x)\tilde u_{{\rm int}\,b}(0))\mid \Omega>_c\eqno (3.5)$$
and the matter propagator by
$$iS_{\alpha\beta}(p)\equiv i\delta_{\alpha\beta}S(p):\=d (2\pi)^{-2}\int
d^4x\,e^{ipx}
<\Omega \mid T(\psi_{{\rm int}\,\alpha}(x)\psq_{{\rm int}\,\beta}(0))\mid
\Omega>_c.\eqno (3.6)$$
Their perturbative expansions read in terms of our $\hat t$-distributions
$$i\tilde D_{ab}(k)={i\over (2\pi)^{2}}{\delta_{ab}\over k^2}+
\delta_{ab}{ik_\mu\over k^2}\sum_{n=1}^\infty
{(2\pi )^{(4n-4)}\over (2n-2)!}
t_{u\tilde  u}^{\mu\,(2n)}(-k,k,0,...,0){1\over k^2},\>\>\>\>\>k^2\not=0,\eqno
(3.7)$$
$$iS_{\alpha\beta}(p)={i\over (2\pi)^{2}}\delta_{\alpha\beta}{\gamma^\nu
p_\nu+m\over p^2-m^2}-$$
$$-\delta_{\alpha\beta}{\gamma^\nu p_\nu+m\over p^2-m^2}\sum_{n=1}^\infty
{(2\pi )^{(4n-4)}\over (2n-2)!}
\tau_{\psq\psi}^{(2n)}(p,-p,0,...,0){\gamma^\nu p_\nu+m\over
p^2-m^2},\>\>\>\>\>\>
p^2\not= m^2.\eqno (3.8)$$
Note that $\tilde D(k)$ (3.5) has the structure
$$\tilde D(k)=\tilde D_0(k)+k_\mu \tilde D^\mu_1(k),\quad\quad {\rm where}\quad
\tilde D_0(k)\=d {1\over (2\pi)^{2}}{1\over k^2}\eqno(3.9)$$
is the contribution of zeroth order.
The factor $k_\mu$ in $k_\mu \tilde D^\mu_1(k)$ originates from the derivative
$\d_\mu$ in $t_{u\tilde  u}^{\mu}:u\d_\mu\tilde u:$. The covariant
decomposition
of $\tilde D^\mu_1(k)$ (3.9) is
$$\tilde D^\mu_1(k)=k^\mu \tilde D_2(k).\eqno(3.10)$$

Now we will consider the 3-point Green's functions corresponding to the
couplings $T_1^A$,
$T_1^u$ and $T_1^\psi$. We are going to define the connected Green's functions
$\tilde T,\,
\tilde G,\,\tilde\Gamma$ and the proper (or one-particle irreducible) vertex
functions
$T,\,G,\,\Gamma$ which are the connected Green's functions with amputated legs
$$igT_{abc}^{\mu'\nu'\tau'}(p,q,k)iD_{\>\>\mu'}^{\mu}(p)iD_{\>\>\nu'}^{\nu}(q)
iD_{\>\>\tau'}^{\tau}(k)\vert_{p+q+k=0}:\=d \tilde T_{abc}^{\mu\nu\tau}(p,q,k)
\vert_{p+q+k=0}:\=d $$
$$\=d (2\pi)^{-4}\int d^4x\int d^4y\>e^{ipx}e^{iqy}<\Omega \mid T(A_{{\rm
int}\,a}^\mu
(x)A_{{\rm int}\,b}^\nu (y)A_{{\rm int}\,c}^\tau (0))\mid
\Omega>_c,\eqno(3.11)$$
$$igG_{abc}^{\mu'}(k,q,p)iD_{\>\>\mu'}^{\mu}(k)i\tilde D(q)i\tilde
D(p)\vert_{p+q+k=0}:\=d
\tilde G_{abc}^{\mu}(k,q,p)\vert_{p+q+k=0}:\=d$$
$$\=d (2\pi)^{-4}\int d^4x\int d^4y\>e^{ipx}e^{iqy}<\Omega \mid T(u_{{\rm
int}\,c}
(x)\tilde u_{{\rm int}\,b}(y)A_{{\rm int}\,a}^\mu (0))\mid
\Omega>_c\eqno(3.12)$$
and
$$iS(p)ig\Gamma_{\alpha\beta\,a}^{\mu'}(p,q,k)iS(-q)iD_{\>\>\mu'}^{\mu}(k)
\vert_{p+q+k=0}:\=d
\tilde \Gamma_{\alpha\beta\,a}^{\mu}(p,q,k)\vert_{p+q+k=0}:\=d$$
$$\=d (2\pi)^{-4}\int d^4x\int d^4y\>e^{ipx}e^{iqy}<\Omega \mid T(\psi_{{\rm
int}\,\alpha}
(x)\psq_{{\rm int}\,\beta}(y)A_{{\rm int}\,a}^\mu (0))\mid
\Omega>_c.\eqno(3.13)$$
By means of the Gell-Mann Low series we can express the perturbative expansions
of $\tilde T,\,
\tilde\Gamma$ and $\tilde G$ in terms of our $t$-distributions. The results in
$n$-th
order ($n\geq 3$) contain non-degenerate terms $\sim\tilde t^{(n)}_{B_1B_2B_3}$
and
degenerate ones which consist of a two-legs distribution
in order $(n-1)$ and one further vertex (fig.2).
We only consider the region $p+q+k=0,\>p^2\not= 0$ rsp. $p^2\not=
m^2,\,q^2\not= 0$
rsp. $q^2\not= m^2,\,k^2\not= 0$
$$\tilde
T_{abc}^{(n)\mu\nu\tau}(p,q,k)\vert_{p+q+k=0}={(2\pi)^{(2n-6)}if_{abc}\over
(n-3)!p^2q^2k^2}
\Bigl\{\tilde t^{(n)\mu\nu\tau}_{AAA}(p,q,k,0,...)+$$
$$+{ig\over (2\pi)^2}\bigl [\tilde t^{(n-1)\mu\mu '}_{AA}(p,-p,0,...){1\over
p^2}
[g_{\mu '}^{\>\>\nu}(p-q)^\tau +g^{\nu\tau}(q-k)_{\mu'}+g_{\mu
'}^{\>\>\tau}(k-p)^\nu]+$$
$$+\tilde t^{(n-1)\nu\nu'}_{AA}(q,-q,0,...){1\over q^2}
[g_{\nu '}^{\>\>\mu}(p-q)^\tau
+g_{\nu'}^{\>\>\tau}(q-k)^\mu+g^{\mu\tau}(k-p)_{\nu'}]+$$
$$+\tilde t^{(n-1)\tau\tau'}_{AA}(k,-k,0,...){1\over k^2}
[g^{\mu\nu}(p-q)_{\tau'}+g_{\tau'}^{\>\>\nu}(q-k)^\mu+g^{\>\>\mu}_{\tau'}
(k-p)^\nu]
\bigr ]\Bigr\}\vert_{p+q+k=0},\eqno(3.14)$$
$$\tilde
\Gamma_{\alpha\beta\,a}^{(n)\mu}(p,q,k)\vert_{p+q+k=0}={(2\pi)^{(2n-6)}i\over
(n-3)!k^2}
(\lambda_a)_{\alpha\beta}{\gamma^\nu p_\nu+m\over
p^2-m^2}\Bigl\{\tilde\tau^{(n)\mu}_{\psq\psi A}
(p,q,k,0,...)+$$
$$+{g\over 2(2\pi)^2}\bigl [\gamma_\lambda {1\over k^2}\tilde
t^{(n-1)\mu\lambda}_{AA}
(k,-k,0,...)-\gamma^\mu {-\gamma^\nu q_\nu+m\over
q^2-m^2}\tau^{(n-1)}_{\psq\psi}(-q,q,0,...)-$$
$$-\tau^{(n-1)}_{\psq\psi}(p,-p,0,...){\gamma^\nu p_\nu+m\over
p^2-m^2}\gamma^\mu
\bigr ]\Bigr\}{-\gamma^\nu q_\nu+m\over q^2-m^2}\vert_{p+q+k=0}\eqno(3.15)$$
and
$$\tilde G^\mu_{abc}(k,q,p)\vert_{p+q+k=0}=\tilde
G^{\mu\lambda}_{1\,abc}(k,q,p)p_\lambda
\vert_{p+q+k=0}\eqno(3.16)$$
with
$$\tilde
G^{(n)\mu\lambda}_{1\,abc}(k,q,p)\vert_{p+q+k=0}={(2\pi)^{(2n-6)}f_{abc}\over
(n-3)!
p^2q^2k^2} \Bigl\{\tilde t^{(n)\mu\lambda}_{Au\tilde u}(k,q,p,0,...)+$$
$$+{g\over (2\pi)^2}\bigl [-\tilde t^{(n-1)\mu\lambda}_{AA}(k,-k,0,...){1\over
k^2}
+g^{\mu\lambda}{iq_\tau\over q^2}t^\tau_{u\tilde u}(q,-q,0,...)-
t^\lambda_{u\tilde u}(-p,p,0,...){ip^\mu\over p^2}\bigr
]\Bigr\}\vert_{p+q+k=0}.\eqno(3.17)$$
Similar to (3.9) the factor $p_\lambda$ in (3.16) comes from $\d_\lambda \tilde
u$ which
is contracted with the external vertex $u_c(x)$ in $<\Omega \mid T(u_{{\rm
int}\,c}
(x)...)\mid \Omega>_c$. Note that the proper vertex $G^\mu_{abc}(k,q,p)$
(3.12) has an external leg $\d_\nu \tilde u$ and that the corresponding factor
$p_\nu$
is absorbed in $G^\mu$
$$G^\mu_{abc}(k,q,p)\vert_{p+q+k=0}=G^{\mu\nu}_{1\,abc}(k,q,p)p_\nu\vert_
{p+q+k=0}.\eqno(3.18)$$
Now we insert (3.9-10) into $\tilde D(p)$ in (3.12). Moreover we use (3.16),
(3.18) there
and omit the factor $p_\lambda$ coming from the contraction of $\d_\lambda
\tilde u$
with the external vertex $u_c(x)$. It results
$$\tilde G^{\mu\lambda}_{1\,abc}(k,q,p)\vert_{p+q+k=0}=
gG_{1\,abc}^{\mu'\nu}(k,q,p)D_{\>\>\mu'}^{\mu}(k)\tilde D(q)
(\tilde D_0(p)g_\nu^{\>\>\lambda}+p_\nu p^\lambda \tilde
D_2(p))\vert_{p+q+k=0}.\eqno(3.19)$$

Let us consider the Slavnov-Taylor identity (formula (1.1) of [19] or (2.8) of
[17] in
the Feynman gauge)
$${p_{1\alpha}\over
p_1^2}T^{\alpha\mu'\nu'}_{abc}(p_1,p_2,p_3)\vert_{p_1+p_2+p_3=0}=
(D^{\nu'\,-1}_{\>\>\lambda} (p_3)+(2\pi)^2p_3^{\nu'}p_{3\lambda})
G_{1\,bac}^{\mu'\lambda}(p_2,p_1,p_3)\tilde D(p_1)+$$
$$+(D^{\mu'\,-1}_{\>\>\lambda} (p_2)+(2\pi)^2p_2^{\mu'}p_{2\lambda})
G_{1\,cab}^{\nu'\lambda}(p_3,p_1,p_2)\tilde
D(p_1)\vert_{p_1+p_2+p_3=0},\eqno(3.20)$$
which is written in terms of one-particle irreducible Green's functions and
$p_i^2\not=0,\>i=1,2,3$
is assumed. We multiply (3.20) with
$D_{\>\>\mu'}^{\mu}(p_2)D_{\>\>\nu'}^{\nu}(p_3)$ and use (3.4). It results
the equivalent equation
$${p_{1\alpha}\over
p_1^2}T^{\alpha\mu'\nu'}_{abc}(p_1,p_2,p_3)D_{\>\>\mu'}^{\mu}(p_2)
D_{\>\>\nu'}^{\nu}(p_3)=D_{\>\>\mu'}^{\mu}(p_2){1\over
p_3^2}(g^\nu_{\>\>\lambda}p_3^2-p_3^\nu
p_{3\lambda})G_{1\,bac}^{\mu'\lambda}(p_2,p_1,p_3)\tilde D(p_1)+$$
$$+D_{\>\>\nu'}^{\nu}(p_3){1\over p_2^2}(g^\mu_{\>\>\lambda}p_2^2-p_2^\mu
p_{2\lambda})G_{1\,cab}^{\nu'\lambda}(p_3,p_1,p_2)\tilde D(p_1).\eqno(3.21)$$
By means of (3.4) the l.h.s. can be expressed by $\tilde T$ (3.11). With (3.19)
the
first term on the r.h.s. reads in terms of $\tilde G$ (3.12) as follows
$$G_{1\,bac}^{\mu'\lambda}(p_2,p_1,p_3)D_{\>\>\mu'}^{\mu}(p_2)\tilde
D(p_1)\tilde D_0(p_3)
(g^\nu_{\>\>\lambda}p_3^2-p_3^\nu p_{3\lambda})={1\over g}\tilde
G_{1\,bac}^{\mu\lambda}(p_2,p_1,p_3)(g^\nu_{\>\>\lambda}p_3^2-p_3^\nu
p_{3\lambda}).\eqno(3.22)$$
Summing up the Slavnov-Taylor identity (3.20) is equivalent to
$$-p_{1\alpha}\tilde
T^{\alpha\mu\nu}_{abc}(p_1,p_2,p_3)\vert_{p_1+p_2+p_3=0}=$$
$$=\tilde G_{1\,bac}^{\mu\lambda}(p_2,p_1,p_3)(g^\nu_{\>\>\lambda}p_3^2-p_3^\nu
p_{3\lambda})
+\tilde G_{1\,cab}^{\nu\lambda}(p_3,p_1,p_2)(g^\mu_{\>\>\lambda}p_2^2-p_2^\mu
p_{2\lambda})
\vert_{p_1+p_2+p_3=0},\eqno(3.23)$$
which is an identity between the connected Green's functions.
(Slavnov's identities [11] are originally written
in terms of {\it connected} Green's functions. For someone who is very familiar
with Slavnov's
formalism it may therefore be easier to write Slavnov's identity {\it directly}
in the form (3.23),
without the detour (3.20) with the one-particle irreducible Green's functions.
(3.20) rsp. (3.23) cannot be
obtained from 't Hooft's diagramatic Ward identities [12] or from Taylor's
identities [10],
because in these identities some of the external momenta are on-shell.)

Now we are going to show that the {\it perturbative version of (3.23) is
essentially our
reduced Cg-identity (2.14)}. For this purpose we insert the $n$-th order
expressions
(3.14) and (3.17). By means of the 2-legs Cg-identity (2.5) we obtain for the
l.h.s. of (3.23)
$${\rm l.h.s.}^{(n)}={(2\pi)^{(2n-6)}f_{abc}\over (n-3)!p_1^2p_2^2p_3^2}
\Bigl\{-ip_{1\alpha}\tilde t^{\alpha\mu\nu}_{AAA}(p_1,p_2,p_3,0,...)+$$
$$+{g\over (2\pi)^2}\bigl [\tilde t^{\mu\mu'}_{AA}(p_2,-p_2,0,...){1\over
p_2^2}
[-p_{1\mu'}p_2^\nu +g_{\mu '}^{\>\>\nu}p_1(p_2-p_3)+p_1^\nu p_{3\mu'}]+$$
$$+\tilde t^{\nu\nu'}_{AA}(p_3,-p_3,0,...){1\over p_3^2}
[-p_1^\mu p_{2\nu'}+g_{\nu '}^{\>\>\mu}p_1(p_2-p_3)+p_{1\nu'}p_3^\mu]
\bigr ]\Bigr\}\vert_{p_1+p_2+p_3=0}=$$
$$={(2\pi)^{(2n-6)}f_{abc}\over (n-3)!p_1^2p_2^2p_3^2}
\Bigl\{-ip_{1\alpha}\tilde t^{\alpha\mu\nu}_{AAA}(p_1,p_2,p_3,0,...)+$$
$$+{g\over (2\pi)^2}\bigl [\tilde t^{\mu\nu}_{AA}(-p_3,p_3,0,...)
-\tilde t^{\nu\mu}_{AA}(-p_2,p_2,0,...)+$$
$$+\tilde t^{\mu\mu'}_{AA}(p_2,-p_2,0,...){1\over
p_2^2}(g_{\mu'}^{\>\>\nu}p_3^2-p_{3\mu'}p_3^\nu)
-\tilde t^{\nu\nu'}_{AA}(p_3,-p_3,0,...){1\over
p_3^2}(g_{\nu'}^{\>\>\mu}p_2^2-p_{2\nu'}p_2^\mu)
\bigr ]\Bigr\}\vert_{p_1+p_2+p_3=0}.\eqno(3.24)$$
Using the covariant decomposition (2.4) of $t^\mu_{u\tilde u}$, the r.h.s. of
(3.23) reads
$${\rm r.h.s.}^{(n)}={(2\pi)^{(2n-6)}f_{abc}\over (n-3)!p_1^2p_2^2p_3^2}
\Bigl\{-\tilde t^{\mu\lambda}_{Au\tilde u}(p_2,p_1,p_3,0,...)
(g^\nu_{\>\>\lambda}p_3^2-p_3^\nu p_{3\lambda})+$$
$$+\tilde t^{\nu\lambda}_{Au\tilde u}(p_3,p_1,p_2,0,...)
(g^\mu_{\>\>\lambda}p_2^2-p_2^\mu p_{2\lambda})+
{g\over (2\pi)^2}\bigl[i\bar t_{u\tilde u}(p^2_1)[-p^2_3 g^{\mu\nu}+p^\mu_3
p^\nu_3
+p^2_2 g^{\mu\nu}-p^\mu_2 p^\nu_2]+$$
$$+\tilde t^{\mu\mu'}_{AA}(p_2,-p_2,0,...){1\over
p_2^2}(g_{\mu'}^{\>\>\nu}p_3^2-p_{3\mu'}p_3^\nu)
-\tilde t^{\nu\nu'}_{AA}(p_3,-p_3,0,...){1\over
p_3^2}(g_{\nu'}^{\>\>\mu}p_2^2-p_{2\nu'}p_2^\mu)
\bigr ]\Bigr\}\vert_{p_1+p_2+p_3=0}.\eqno(3.25)$$
The equality (3.24)=(3.25) is exactly our reduced Cg-identity (2.14). This
proves the
Slavnov-Taylor identity (3.23) in the framework of causal perturbation theory.

(3.20) rsp. (3.23) is the {\it only} Slavnov-Taylor identity with three
external legs in
pure Yang-Mills theories [10,11]. If we add the coupling to matter fields,
there is precisely
one additional identity of this kind [10],
which corresponds to the reduced Cg-identity (2.9). In order to formulate this
identity, {\it Taylor is forced to introduce the Q-vertex} [10]. (This is in
accordance with
the fact that the Q-vertex cannot be eliminated completely in (2.9).) We define
the connected
Green's functions $S^F(p)\gamma^\mu\tilde\Gamma^1(p,q,k)$
and $\tilde\Gamma^2(p,q,k)\gamma^\mu S^F(-q)$
$$S^F(p)\gamma^\mu\tilde\Gamma_{\alpha\beta\,a}^1(p,q,k)\vert_{p+q+k=0}:\=d
{-g\over 2(2\pi)^4}\int d^4x\int d^4y\>e^{ipx}e^{iqy}\int d^4x_1$$
$$<\Omega \mid
T(S^F(x-x_1)(\lambda_{a'})_{\alpha\alpha'}\gamma^\mu\psi_{{\rm
int}\,\alpha'}(x_1)
u_{{\rm int}\,a'}(x_1)\psq_{{\rm int}\,\beta}(y)\tilde u_{{\rm int}\,a}(0))\mid
\Omega>_c,
\eqno(3.26)$$
$$\tilde\Gamma_{\alpha\beta\,a}^2(p,q,k)\gamma^\mu S^F(-q)\vert_{p+q+k=0}:\=d
{-g\over 2(2\pi)^4}\int d^4x\int d^4y\>e^{ipx}e^{iqy}\int d^4x_1$$
$$<\Omega \mid
T(\psi_{{\rm int}\,\alpha}(x)\psq_{{\rm
int}\,\beta'}(x_1)(\lambda_{a'})_{\beta'\beta}
\gamma^\mu S^F(x_1-y)u_{{\rm int}\,a'}(x_1)\tilde u_{{\rm int}\,a}(0))\mid
\Omega>_c.
\eqno(3.27)$$
They have one Q-vertex at $x_1$ which is directly contracted with $\psi_\alpha
(x)$ rsp.
$\psq_\beta (y)$, all other vertices are ordinary vertices. By means of the
Gell-Mann
Low series we obtain
$$\tilde\Gamma_{\alpha\beta\,a}^{(n)\,1}(p,q,k)\vert_{p+q+k=0}=
{(2\pi)^{(2n-4)}i\over
(n-3)!k^2}
(\lambda_a)_{\alpha\beta}\Bigl\{\bar\tau^{(n)\,1}_{\psq\psi u}(p,q,k,0,...)+$$
$$+{g\over 2(2\pi)^2}\bigl [-it^{(n-1)\tau}_{u\tilde u}(k,-k,0,...){k_\tau\over
k^2}
-{-\gamma^\nu q_\nu+m\over q^2-m^2}\tau^{(n-1)}_{\psq\psi}(-q,q,0,...)
\bigr ]\Bigr\}{-\gamma^\nu q_\nu+m\over q^2-m^2}\vert_{p+q+k=0},\eqno(3.28)$$
$$\tilde\Gamma_{\alpha\beta\,a}^{(n)\,2}(p,q,k)\vert_{p+q+k=0}=
{(2\pi)^{(2n-4)}i\over
(n-3)!k^2}
(\lambda_a)_{\alpha\beta}{\gamma^\nu p_\nu+m\over p^2-m^2}
\Bigl\{\bar\tau^{(n)\,2}_{\psq\psi u}(p,q,k,0,...)+$$
$$+{g\over 2(2\pi)^2}\bigl [-it^{(n-1)\tau}_{u\tilde u}(k,-k,0,...){k_\tau\over
k^2}
-\tau^{(n-1)}_{\psq\psi}(p,-p,0,...){\gamma^\nu p_\nu+m\over p^2-m^2}
\bigr ]\Bigr\}\vert_{p+q+k=0},\eqno(3.29)$$
in $n$-th order perturbation theory ($n\geq 3$) and $p^2\not= m^2,\,
q^2\not= m^2,\,k^2\not= 0$ is assumed. In terms of connected Green's functions
Taylor's identity reads
$$-p_{3\mu}\tilde\Gamma_{\alpha\beta\,a}^\mu
(p_1,p_2,p_3)\vert_{p_1+p_2+p_3=0}=$$
$$=S^F(p_1)(\gamma^\mu p_{1\mu}-m)\tilde\Gamma_{\alpha\beta\,a}^1(p_1,p_2,p_3)+
\tilde\Gamma_{\alpha\beta\,a}^2(p_1,p_2,p_3)(\gamma^\mu p_{2\mu}+m) S^F(-p_2)
\vert_{p_1+p_2+p_3=0}\equiv $$
$$\equiv
(2\pi)^{-2}[\tilde\Gamma_{\alpha\beta\,a}^1(p_1,p_2,p_3)-\tilde\Gamma_
{\alpha\beta\,a}^2
(p_1,p_2,p_3)]\vert_{p_1+p_2+p_3=0},\eqno(3.30)$$
where the external momenta are off-shell. In order to obtain Taylor's original
identity
(formula (14) of [10]) from (3.30), one has to express $\tilde\Gamma$ (3.13),
$\tilde\Gamma^1$,
$\tilde\Gamma^2$ by proper vertices and by 2-point Green's functions, and to
amputate the
three external Feynman propagators. Then, one must multiply by $\bar u(p_1)$,
which
eliminates the $\tilde\Gamma^1$-term on the r.h.s.. This multiplication puts
$p_1$ on the
mass-shell $\bar u(p_1)\gamma^\nu p_{1\nu}=\bar u(p_1) m$ and, therefore,
{\it infrared divergences} appear.

Let us consider (3.30) in perturbation theory. We insert the $n$-th order
expressions (3.15),
(3.28) and (3.29). With the 2-legs Cg-identity (2.5) the l.h.s. of (3.30)
becomes
$${\rm l.h.s.}^{(n)}={(2\pi)^{(2n-6)}i\over
(n-3)!p_3^2}(\lambda_a)_{\alpha\beta}
{\gamma^\nu p_{1\nu}+m\over
p_1^2-m^2}\Bigl\{-p_{3\mu}\tilde\tau^{(n)\mu}_{\psq\psi A}
(p_1,p_2,p_3,0,...)+$$
$$+{g\over 2(2\pi)^2}\bigl [\tau^{(n-1)}_{\psq\psi}(-p_2,p_2,0,...)
-\tau^{(n-1)}_{\psq\psi}(p_1,-p_1,0,...)+$$
$$+(\gamma^\mu (p_3+p_2)_\mu+m) {-\gamma^\nu p_{2\nu}+m\over p_2^2-m^2}
\tau^{(n-1)}_{\psq\psi}(-p_2,p_2,0,...)+$$
$$+\tau^{(n-1)}_{\psq\psi}(p_1,-p_1,0,...){\gamma^\nu p_{1\nu}+m\over
p_1^2-m^2}
(\gamma^\mu (p_3+p_1)_\mu-m)
\bigr ]\Bigr\}{-\gamma^\nu p_{2\nu}+m\over
p_2^2-m^2}\vert_{p_1+p_2+p_3=0}.\eqno(3.31)$$
Using the covariant decomposition (2.4) of $t^\mu_{u\tilde u}$, we obtain for
the r.h.s.
$${\rm r.h.s.}^{(n)}={(2\pi)^{(2n-6)}i\over
(n-3)!p_3^2}(\lambda_a)_{\alpha\beta}
{\gamma^\nu p_{1\nu}+m\over p_1^2-m^2}\Bigl\{(\gamma^\mu p_{1\mu}-m)\bar
\tau^{(n)1}_{\psq\psi u}(p_1,p_2,p_3,0,...)+$$
$$+\bar\tau^{(n)\,2}_{\psq\psi u}(p_1,p_2,p_3,0,...)(\gamma^\mu p_{2\mu}+m)
+{g\over 2(2\pi)^2}\bigl [i\gamma_\mu t^{(n-1)\mu}_{u\tilde
u}(p_3,-p_3,0,...)+$$
$$+(-\gamma^\mu p_{1\mu}+m) {-\gamma^\nu p_{2\nu}+m\over p_2^2-m^2}
\tau^{(n-1)}_{\psq\psi}(-p_2,p_2,0,...)-$$
$$-\tau^{(n-1)}_{\psq\psi}(p_1,-p_1,0,...){\gamma^\nu p_{1\nu}+m\over
p_1^2-m^2}
(\gamma^\mu p_{2\mu}+m)
\bigr ]\Bigr\}{-\gamma^\nu p_{2\nu}+m\over
p_2^2-m^2}\vert_{p_1+p_2+p_3=0}.\eqno(3.32)$$
The equality (3.31)=(3.32) is precisely the identity (2.9).
{\it The reduced Cg-identities (2.5) and (2.9) imply the perturbative version
of the Slavnov-Taylor identity (3.30)}.
\vskip 0.5cm
{\it (c) Four-Legs Slavnov-Taylor Identities}
\vskip 0.5cm
We define the connected 4-point Green's functions
$$\tilde L_{abcd}^{\alpha\mu\nu\tau}(p,q,k,r)\vert_{p+q+k+r=0}:\=d
(2\pi)^{-6}\int d^4x\int d^4y\int d^4z\>e^{ipx}e^{iqy}e^{ikz}$$
$$<\Omega \mid T(A_{{\rm int}\,a}
^\alpha (x)A_{{\rm int}\,b}^\mu (y)A_{{\rm int}\,c}^\nu (z)A_{{\rm
int}\,d}^\tau (0))
\mid \Omega>_c,\eqno(3.33)$$
$$\tilde H_{abcd}^{\lambda\mu\nu}(p,q,k,r)q_\lambda\vert_{p+q+k+r=0}:\=d
(2\pi)^{-6}\int d^4x\int d^4y\int d^4z\>e^{ipx}e^{iqy}e^{ikz}$$
$$<\Omega \mid
T(\tilde u_{{\rm int}\,a}(x) u_{{\rm int}\,b}(y)A_{{\rm int}\,c}^\mu (z)A_{{\rm
int}\,d}
^\nu (0))\mid \Omega>_c,\eqno(3.34)$$
$$\tilde M_{\alpha\beta\,ab}^{\mu\nu}(p,q,k,r)\vert_{p+q+k+r=0}:\=d
(2\pi)^{-6}\int d^4x\int d^4y\int d^4z\>e^{ipx}e^{iqy}e^{ikz}$$
$$<\Omega \mid
T(\psi_{{\rm int}\,\alpha}(x)\psq_{{\rm int}\,\beta}(y)A_{{\rm int}\,a}^\mu
(z)A_{{\rm int}\,b}
^\nu (0))\mid \Omega>_c,\eqno(3.35)$$
$$\tilde F_{\alpha\beta\,ab}^{\lambda}(p,q,k,r)r_\lambda\vert_{p+q+k+r=0}:\=d
(2\pi)^{-6}\int d^4x\int d^4y\int d^4z\>e^{ipx}e^{iqy}e^{ikz}$$
$$<\Omega \mid
T(\psi_{{\rm int}\,\alpha}(x)\psq_{{\rm int}\,\beta}(y)\tilde u_{{\rm
int}\,a}(z)
u_{{\rm int}\,b}(0))\mid \Omega>_c,\eqno(3.36)$$
$$S^F(p)\gamma^\mu\tilde
M_{\alpha\beta\,ab}^{1\,\nu}(p,q,k,r)\vert_{p+q+k+r=0}:\=d
{-g\over 2(2\pi)^6}\int d^4x\int d^4y\int d^4z\>e^{ipx}e^{iqy}e^{ikz}\int
d^4x_1$$
$$<\Omega \mid T(S^F(x-x_1)(\lambda_{a'})_{\alpha\alpha'}\gamma^\mu\psi_{{\rm
int}\,\alpha'}(x_1)
u_{{\rm int}\,a'}(x_1)\psq_{{\rm int}\,\beta}(y)\tilde u_{{\rm int}\,a}(z))
A_{{\rm int}\,b}^\nu (0))\mid \Omega>_c,\eqno(3.37)$$
$$\tilde M_{\alpha\beta\,ab}^{2\,\nu}(p,q,k,r)\gamma^\mu
S^F(-q)\vert_{p+q+k+r=0}:\=d
{-g\over 2(2\pi)^6}\int d^4x\int d^4y\int d^4z\>e^{ipx}e^{iqy}e^{ikz}\int
d^4x_1$$
$$<\Omega \mid T(\psi_{{\rm int}\,\alpha}(x)\psq_{{\rm
int}\,\beta'}(x_1)(\lambda_{a'})_{\beta'\beta}
\gamma^\mu S^F(x_1-y)u_{{\rm int}\,a'}(x_1)\tilde u_{{\rm int}\,a}(z)
A_{{\rm int}\,b}^\nu (0))\mid \Omega>_c.\eqno(3.38)$$
The latter two Green's functions have one Q-vertex at $x_1$, all other vertices
are ordinary
vertices. In (3.34), (3.36) a factor $q_\lambda$ (rsp. $r_\lambda$) is
separated. Similarly to
(3.9), (3.16), this factor comes from $\d_\lambda \tilde u$ which
is contracted with the external vertex $u_b(y)$ (rsp. $u_b(0)$) in $<\Omega
\mid T(...
u_{{\rm int}\,b}(...)...)\mid \Omega>_c$. Although 't Hooft [12] and Taylor
[10] consider
only the case with some of the external momenta on the mass-shell, it is clear
from their works
how to translate the three-legs Slavnov-Taylor identities (3.23), (3.30) to the
four-legs case:
Assuming all external momenta $p_i\>(i\in\{1,2,3,4\})$ and all $(p_i+p_j)\>
(1\leq i<j\leq 4)$ to be off-shell, one has
$$p_{1\alpha}\tilde
L^{\alpha\mu\nu\tau}_{abcd}(p_1,p_2,p_3,p_4)\vert_{p_1+p_2+p_3+p_4=0}=
\tilde
H_{abcd}^{\lambda\nu\tau}(p_1,p_2,p_3,p_4)(g^\mu_{\>\>\lambda}p_2^2-p_2^\mu
p_{2\lambda})+$$
$$+\tilde
H_{acdb}^{\lambda\tau\mu}(p_1,p_3,p_4,p_2)(g^\nu_{\>\>\lambda}p_3^2-p_3^\nu
p_{3\lambda})
+\tilde
H_{adbc}^{\lambda\mu\nu}(p_1,p_4,p_2,p_3)(g^\tau_{\>\>\lambda}p_4^2-p_4^\tau
p_{4\lambda})
\vert_{p_1+p_2+p_3+p_4=0},\eqno(3.39)$$
$$-p_{3\mu}\tilde
M_{\alpha\beta\,ab}^{\mu\nu}(p_1,p_2,p_3,p_4)\vert_{p_1+p_2+p_3+p_4=0}=
S^F(p_1)(\gamma^\mu p_{1\mu}-m)\tilde
M_{\alpha\beta\,ab}^{1\,\nu}(p_1,p_2,p_3,p_4)+$$
$$+\tilde M_{\alpha\beta\,ab}^{2\,\nu}(p_1,p_2,p_3,p_4)(\gamma^\mu p_{2\mu}+m)
S^F(-p_2)-
\tilde
F_{\alpha\beta\,ab}^{\lambda}(p_1,p_2,p_3,p_4)(g^\nu_{\>\>\lambda}p_4^2-p_4^\nu
p_{4\lambda})
\vert_{p_1+...=0}\equiv $$
$$\equiv (2\pi)^{-2}[\tilde M_{\alpha\beta\,ab}^{1\,\nu}(p_1,p_2,p_3,p_4)-
\tilde M_{\alpha\beta\,ab}^{2\,\nu}(p_1,p_2,p_3,p_4)]-$$
$$-\tilde
F_{\alpha\beta\,ab}^{\lambda}(p_1,p_2,p_3,p_4)(g^\nu_{\>\>\lambda}p_4^2-p_4^\nu
p_{4\lambda})\vert_{p_1+p_2+p_3+p_4=0}\eqno(3.40)$$
and these are {\it all} Slavnov-Taylor identities with four external legs. The
identity (3.39) in
terms of one-particle irreducible Green's functions can be found in [19],
formula (1.2).

We claim that {\it the perturbative versions of the Slavnov-Taylor identities
(3.39-40) can
be proven from our reduced two- ,three- and four-legs Cg-identities}. Since the
proof of
this statement is a straight-forward calculation, completely analogous to the
three-legs
case, but very much longer, we only describe the procedure here.
First one has to express the Gell-Mann Low series of the connected
Green's functions (3.33-38) in terms of our $t$-distributions. Besides the
non-degenerate
terms $\sim\tilde t^{(n)}_{B_1B_2B_3B_4}$ of order $n$
there are degenerate ones, namely three-legs distributions
in order $(n-1)$ with one separated vertex (fig.3a), and two-legs distributions
in order $(n-2)$
with two separated vertices (figs.3b,c). In the case of $\tilde L$ (3.33) one
should not forget
the two-legs distribution $\tilde t^{(n-2)}_{AA}$ combined with a four-gluon
vertex (fig.3d). All
other terms with four-gluon vertices (e.g. fig.1) are contained in the $\tilde
t$-distributions.
Next we insert these $n$-th order expressions for the connected Green's
functions into (3.39-40).
In order to obtain agreement of the l.h. and r.h. sides in the resulting
equations one uses

-the reduced Cg-identities (2.24), (2.14), (2.5), the covariant
decomposition (2.4) of $t_{u\tilde u}$ and the Jacobi-identity of the structure
constants $f_{abc}$
in the case of (3.39);

-and the reduced Cg-identities (2.23), (2.9), (2.14), (2.5), again (2.4)
and the identity $[\lambda_a,\lambda_b]=2if_{abc}\lambda_c$ in the case of
(3.40).
\vskip 0.5cm
{\it (d) Concluding Remarks}
\vskip 0.5cm
There are also Cg- and Slavnov-Taylor identities with five or more external
legs. The crucial
step in the proof of gauge invariance is the distribution splitting. In this
process the
mentioned Cg-identities cannot be violated [4,5]. Consequently, they give no
further restriction
on the normalization of the $t$-distributions (sect.2d). Therefore,
they are not of great interest and we do not consider them here.

The original Cg-identities [2,3,5] contain more information than the
Slavnov-Taylor identities,
because their coordinates refer to external and to {\it inner} vertices,
whereas in the Green's
functions the inner vertices are integrated out with $g(x)\equiv 1$. In
momentum space,
this integration corresponds to the limit of vanishing inner momenta and
vanishing $p_n\=d -(p_1+p_2+...+p_{n-1})$. In sect.2 we have seen that the
unphysical Q-vertex
can only be eliminated from the Cg-identities by taking this limit. The result
of this
elimination are seven reduced Cg-identities. Let us first consider only five of
them:
(2.5) with two legs, (2.9) and (2.14)
with three legs, and (2.23-24) with four legs. We have proven that these five
identities imply the perturbative versions of all Slavnov-Taylor identities up
to
four external legs, namely (3.4) with two legs, (3.23) and (3.30) with three
legs,
and (3.39-40) with four legs. However, our calculations yield more
information:
In the framework of perturbation theory these five Slavnov-Taylor identities
are
{\it equivalent} to the above five reduced Cg-identities.

We turn to the remaining two reduced Cg-identities: (2.15) with three legs and
(2.25)
with four legs. Most probably, they can be expressed by two identities between
connected
Green's functions. The connected Green's function with non-degenerate part
$\sim\bar
t^3_{uu\tilde u}$ (rsp. $\sim\tilde t^{3\nu}_{uu\tilde uA}$), the latter
appears in (2.15)
(rsp. (2.25)), has one Q-vertex $T^u_{1/1}$ (1.11), which is directly
contracted with the
external vertex $u(z)$ in $<\Omega \mid T(\tilde u_{\rm int}(x)\tilde u_{\rm
int}(y)
u_{\rm int}(z)[A_{\rm int}(0)])\mid \Omega>_c$. This is analogous to
$\tilde\Gamma^{1/2}$
(3.26-27) or $\tilde M^{1/2}$ (3.37-38). However, the {\it reduced
Cg-identities (2.15),
(2.25) give no restriction on the normalization of
the physical theory}, in contrast to (2.9), (2.23),
which contain distributions with one Q-vertex, too (sect.2d). Therefore, we do
not consider
the identities of Green's functions corresponding to (2.15), (2.25). They
cannot be found
in the literature either.
\vskip 1cm
{\trm Appendix: Epstein and Glaser's definition\break\vskip 0.3cm of Green's
functions}
\vskip 1cm
Following Epstein and Glaser
(sect.8.1 of [6]) we define a bigger theory by giving its first order
$$S_1(g,g_1,...g_l)\=d \int d^4
x\{T_1(x)g(x)+i\Phi_1(x)g_1(x)+...+i\Phi_l(x)g_l(x)\}.\eqno(A.1)$$
The interacting field $\Phi_{j\,{\rm int}}(x;g)$ is defined by [20,6,21]
$$\Phi_{j\,{\rm int}}(x;g)\=d S^{-1}(g,0,...0){\delta S(g,g_1,...g_l)\over
i\delta g_j(x)}
\Bigl\vert_{g_1=...g_l=0},\>\>\>\>j\in\{1,...l\}. \eqno(A.2)$$
Higher functional derivatives
$$\hat T_{j_1...j_r}(x_1,...x_r;g)\=d S^{-1}(g,0,...0){\delta^r
S(g,g_1,...g_l)\over
i^r\delta g_{j_1}(x_1)...\delta
g_{j_r}(x_r)}\Bigl\vert_{g_1=...g_l=0},\eqno(A.3)$$
$j_1,...j_r\in\{1,...l\}$, define time-ordered products of these interacting
fields [6]
$$\hat T_{j_1...j_r}(x_1,...x_r;g)=\Phi_{j_{\pi 1}\,{\rm int}}(x_{\pi 1};g)
\Phi_{j_{\pi 2}\,{\rm int}}(x_{\pi 2};g)...\Phi_{j_{\pi r}\,{\rm int}}(x_{\pi
r};g),\eqno(A.4)$$
where $\pi\in{\cal S}_r$ is a permutation which puts the coordinates in
time-order
$x_{\pi 1}\succeq x_{\pi 2}\succeq ...x_{\pi r}$ ($x\succeq y$ means
$x\in {\bf R}^4\setminus (y+\bar V^-)$).
One easily obtains the perturbative expansion of (A.3)
$$\hat T_{j_1...j_r}(x_1,...x_r;g)=\sum_{n=0}^{\infty}{1\over i^r n!}
\int
d^4y_1...d^4y_n\,A_{0...0j_1...j_r}(y_1,...y_n;x_1,...x_r)g(y_1)...g(y_n),
\eqno(A.5)$$
with
$$A_{0...0j_1...j_r}(y_1,...y_n;x_1,...x_r)\=d \sum_{I\subset Y}\bar
T_{0...0}(I)
T_{0...0j_1...j_r}(Y\setminus I;x_1,...x_r),\eqno(A.6)$$
where $Y\=d \{y_1,...y_n\}$ and a lower index $0$ means that the vertex at the
corresponding
position in the argument is an ordinary vertex $T_1(x)$,
an index $j$ indicates a vertex $i\Phi_j (x)$. By means of the causal
factorization of
$\bar T_{0...0}(I)$ and $T_{0...0j_1...j_r}(Y\setminus I;x_1,...x_r)$ one
easily proves
$${\rm
supp}\>A_{0...0j_1...j_r}(y_1,...y_n;x_1,...x_r)\subset\{(y_1,...y_n;
x_1,...x_r)\mid
y_i\in\{x_1,...x_r\}+\bar V^-,\,\forall i=1,...n\}.\eqno(A.7)$$
The time-ordering (A.4) relies on this support property (see [6]).
The r-point Green's function (corresponding to (3.1)) is defined [6] by the
vacuum
expectation value of (A.3)
$$<0\mid\hat T_{j_1...j_r}(x_1,...x_r;g)\mid 0>=$$
$$=\sum_{n=0}^{\infty}{1\over i^r n!}
\int d^4y_1...d^4y_n<0\mid A_{0...0j_1...j_r}(y_1,...y_n;x_1,...x_r)\mid
0>g(y_1)...g(y_n)\eqno(A.8)$$
in the Fock vacuum $\mid 0>$ of {\it free} fields. This is in contrast to
(3.1),
where the interacting vacuum $\mid\Omega>$ is used. However, (A.5) is an
expansion
of $T(\Phi_{j_1\,{\rm int}}(x_1)...\Phi_{j_r\,{\rm int}}(x_r))$ in terms of
{\it free} fields,
it is an operator in the Fock space of {\it free} fields, consequently the {\it
free}
vacuum must be used.

The adiabatic switching with $g(y_1)...g(y_n),\>g\in{\cal S}({\bf R}^4)$ in
(A.8)
is an infrared regularization, which should finally be removed. The {\it
crucial
question is whether the
adiabatic limit $g\rightarrow 1$ of (A.8) exists in the sense of tempered
distributions in
$(x_1,...x_r)\in {\bf R}^{4r}$, if a suitable normalization is chosen} (see
(3.3a) and the
remark at the end of this appendix). We shall meet an explicit example where
this limit exists, although the Gell-Mann Low expression does not.
For massive theories Epstein and Glaser (sect.8.2 of [6]) prove the existence
of the
adiabatic limit of (A.8). Moreover they show that this limit possesses all the
{\it linear} properties
of a Green's function such as translation invariance, Lorentz covariance,
causality
and the spectral condition. Blanchard and Seneor [22] prove similar results for
theories
with (some) massless particles such as QED and \break $\lambda :\Phi^{2n}:$
theories. However,
in our Yang-Mills model with matter fields the infrared behaviour is worse
([23] or compare (3.3a) and (3.3b)) and I do not know whether the adiabatic
limit of
(A.8) exists. But our assumption (1.24-25) (extended to the retarded $\hat r$-
and
advanced $\hat a$-distributions [6,8]) implies its existence,
if all (sums of) external momenta are off-shell. This relies on the fact that
the overlapping singularities (appearing in the products of propagators and
non-degenerate
$\hat t,\hat r,\hat a$-distributions) are excluded in this region.

In order to compare the definition (A.8) of Green's functions with the usual
definition
(3.1), we compute (A.8) for the gluon propagator in lowest non-trivial order:
Choosing
$$\Phi_1(x)g_1(x)=A_{a\mu}(x)g^\mu_{1\,a}(x)\eqno(A.9)$$
in (A.1), one obtains
$$<0\mid A_{0011}(y_1,y_2;x_1,x_2)^{\mu\nu}\mid 0>=
C[\tilde t^{\mu\nu}_{AA}(y_1-y_2)D_0^{\rm av}(y_1-x_1)D_0^{\rm av}(y_2-x_2)+$$
$$+\tilde a^{\mu\nu}_{AA}(y_1-y_2)D_0^{+}(y_1-x_1)D_0^{\rm av}(y_2-x_2)
+\tilde r^{\mu\nu}_{AA}(y_1-y_2)D_0^{\rm av}(y_1-x_1)D_0^{+}(y_2-x_2)]
+C[y_1\leftrightarrow y_2],\eqno(A.10)$$
where $\tilde r_{AA}\,(\tilde a_{AA})$ are the retarded (advanced)
distributions [8]
belonging to $\tilde t_{AA}$ (2.1) and
$$D_m^{\rm av}(x)\=d -(2\pi)^{-4}\int d^4k\,{e^{-ikx}\over
k^2-m^2-ik_00},\>\>\>\>\>
D_m^\pm (x)\=d {i\over (2\pi)^3}\int d^4k\,e^{-ikx} \delta(k^2-m^2)\Theta (\pm
k_0).\eqno(A.11)$$
(Note $m=0$ in (A.10).)
C is a constant factor which is not of interest here. Obviously, (A.10) has the
support property (A.7) and it does not agree with the corresponding expression
$$C\tilde
t^{\mu\nu}_{AA}(y_1-y_2)D_0^F(y_1-x_1)D_0^F(y_2-x_2)+(y_1\leftrightarrow
y_2)\eqno(A.12)$$
($D_m^F$ is the Feynman propagator, $D_m^F=D_m^{\rm av}+D_m^+$)
in the Gell-Mann Low series (without integrating over $y_1,y_2$, compare
(3.2-3)).
Let us study the adiabatic limit of (A.10) inserted into (A.8) in momentum
space.
We do this by replacing $g(x)$ by $g_\epsilon (x)$ (1.27) and considering
$\epsilon\rightarrow 0,\,
\epsilon >0$.
Then, with (1.28) we obtain
$$\lim_{\epsilon\to 0}\int d^4x_1d^4x_2\, e^{i(p_1x_1+p_2x_2)}
\int d^4y_1d^4y_2<0\mid A_{0011}(y_1,y_2;x_1,x_2)\mid 0>g_\epsilon
(y_1)g_\epsilon (y_2)\sim$$
$$\sim\lim_{\epsilon\to 0}\delta^4(p_1+p_2)\int d^4k_1d^4k_2\,\hat g_0(k_1)\hat
g_0(k_2)
[\tilde t_{AA}(p_1-\epsilon k_1)\hat D_m^{\rm av}(-p_1)\hat D_m^{\rm
av}(p_1-\epsilon (k_1+k_2))+$$
$$+\tilde a_{AA}(p_1-\epsilon k_1)\hat D_m^{+}(-p_1)\hat D_m^{\rm
av}(p_1-\epsilon (k_1+k_2))
+\tilde r_{AA}(p_1-\epsilon k_1)\hat D_m^{\rm av}(-p_1)\hat
D_m^{+}(p_1-\epsilon (k_1+k_2))]=$$
$$=\lim_{\epsilon\to 0}\delta^4(p_1+p_2)\int d^4k_1d^4k_2\,\hat g_0(k_1)\hat
g_0(k_2)
[\tilde t_{AA}(p_1-\epsilon k_1)\hat D_m^F(p_1)\hat D_m^F(p_1-\epsilon
(k_1+k_2))-\eqno(A.13a)$$
$$+(\tilde t_{AA}(p_1-\epsilon k_1)+\tilde a'_{AA}(p_1-\epsilon k_1)+
\tilde r'_{AA}(p_1-\epsilon k_1))\hat D_m^{-}(p_1)\hat D_m^+(p_1-\epsilon
(k_1+k_2))+$$
$$-\tilde a'_{AA}(p_1-\epsilon k_1)\hat D_m^{-}(p_1)\hat D_m^F(p_1-\epsilon
(k_1+k_2))
+\tilde r'_{AA}(p_1-\epsilon k_1)\hat D_m^F(p_1)\hat D_m^+(p_1-\epsilon
(k_1+k_2))],\eqno(A.13b)$$
where $m=0$.
$\tilde a'_{AA}=\tilde a_{AA}-\tilde t_{AA}$ and $\tilde r'_{AA}=\tilde
r_{AA}-\tilde t_{AA}$
are the usual $a',\,r'$-distributions, which have the factors [2]
$$\tilde a'_{AA}(k)\sim\Theta (k^2-m_1^2)\Theta (k^0),\>\>\>\>\>
\tilde r'_{AA}(k)\sim\Theta (k^2-m_1^2)\Theta (-k^0).\eqno(A.14)$$
(More precisely, $\tilde a'_{AA},\,\tilde r'_{AA}$ are the sum of the gluon,
ghost and matter
loops. For the first two we have $m_1=0$ and for the latter $m_1=2m_\psi\geq
0$.) For
simplicity we assume $\hat g_0$ to have compact support. We first consider a
{\it massive
theory}: $m,\,m_1>0$. Then, the adiabatic limit in (A.13) exists in the sense
of tempered
distributions in $(p_1,p_2)\in {\bf R}^8$ [6]. Due to the
support properties of $\hat D_m^{+}$ (A.11) and $\tilde a'_{AA},\,\tilde
r'_{AA}$ (A.14)
the terms in (A.13b) vanish for $\epsilon >0$ sufficently small. The remaining
term (A.13a) is
an infrared-improved version of
the Gell-Mann Low expression (3.3). We now turn to {\it our model}: In order to
avoid discussions
about infrared divergences, we only consider the region $p_1^2\not= 0$. There,
the
adiabatic limit of (A.13a,b) and the Gell-Mann Low expression exist.
Moreover, for $p_1$ off-shell, the terms (A.13b) vanish if
$\epsilon >0$ is sufficently small and the two definitions of the Gluon
propagator
agree in second order. This holds true also for the ghost and matter
propagators (compare (3.5-6))
since they have a similar structure (A.10), (A.12), (A.13).

In order to get a better understanding of the adiabatic limit of (A.8), we
consider
the electron propagator in second order in QED with the mass of the electron
$m>0$.
The Gell-Mann Low expression (see (3.8))
$$iS^{(2)}(p)=-{\gamma^\nu p_\nu+m\over p^2-m^2+i0}
t_{\psq\psi}^{(2)}(p){\gamma^\nu p_\nu+m\over p^2-m^2+i0}\eqno (A.15)$$
does not exist for $p^2\approx m^2$, because in this region
$t_{\psq\psi}^{(2)}$ [8]
has the behaviour
$$t_{\psq\psi}^{(2)}(p)\sim \bigl\{{\rm log}\>(1-{p^2+i0\over
m^2})(m-{\gamma^\nu p_\nu
\over 2})(1-{m^2\over p^2})+C(\gamma^\nu p_\nu -m)\bigr\},\eqno(A.15a)$$
where $C$ is an arbitrary constant and the mass normalization
$t_{\psq\psi}^{(2)}
(\gamma^\nu p_\nu =m)=0$ is done. The Epstein-Glaser expression is obtained
from (A.13a,b)
by replacing $\hat D^b_m(p)\>(b=\pm ,F)$ by $\hat S^b_m(p)\=d \pm (\gamma^\nu
p_\nu +m)
\hat D^b_m(p)$ and $\tilde f_{AA}(p)\>(f=t,a',r')$ by $f_{\psq\psi}(p)$. Note
that
$\hat S^\pm_m(p)$, rsp. $a'_{\psq\psi}(k),\>r'_{\psq\psi}(k)$ have the same
support
properties as $\hat D^\pm_m(p)$ (A.11), rsp. $\tilde a'_{AA}(k),\>\tilde
r'_{AA}(k)$ (A.14)
(the latter with $m_1$ replaced by $m$). Consequently, the terms corresponding
to (A.13b)
vanish in the adiabatic limit and it remains
$$\sim\lim_{\epsilon\to 0}\delta^4(p_1+p_2)\int d^4k_1d^4k_2\,\hat g_0(k_1)\hat
g_0(k_2)
{\gamma^\nu p_{1\nu}+m\over p_1^2-m^2+i0}\cdot$$
$$\cdot t_{\psq\psi}^{(2)}(p_1-\epsilon k_1){\gamma^\nu (p_1-\epsilon
(k_1+k_2))_\nu+m
\over (p_1-\epsilon (k_1+k_2))^2-m^2+i0}.\eqno(A.15b)$$
An explicit calculation shows that this adiabatic limit exists in the sense of
tempered
distributions in $(p_1,p_2)\in {\bf R}^8$: The
critical term is the product of $f(p)\=d {1\over p^2-m^2+i0}=P({1\over
p^2-m^2})-
i\pi\delta (p^2-m^2)$ with $h(p-\epsilon k_1)\=d {\rm log}\>(1-{(p-\epsilon
k_1)^2+i0\over m^2})=
{\rm log}\>\vert 1-{(p-\epsilon k_1)^2\over m^2}\vert -i\pi\Theta ((p-\epsilon
k_1)^2-m^2)$.
Smeared out in $p$ with an arbitrary test-function, the products $({\rm
Re}\>f)({\rm Re}\>h)$
and $({\rm Im}\>f)({\rm Im}\>h)$ are finite for $\epsilon \rightarrow 0$,
whereas
$({\rm Re}\>f)({\rm Im}\>h)$ and $({\rm Im}\>f)({\rm Re}\>h)$ have divergent
terms
$\sim {\rm log}\>\vert \epsilon\vert$. But the latter cancel exactly. This is
in
accordance with the general result of Blanchard and Seneor [22]. Obviously,
the Epstein-Glaser expression (A.15b) is an infrared-improved version of
(A.15).
The existence of Green's functions in perturbation theory does not only depend
on the behaviour
of the specific theory near the mass-shell (e.g. (3.3a,b), (A.15)), it depends
also
on the definition of the product of distributions with overlapping
singularities.
In the following we always assume the Gell-Mann Low expressions to be
infrared-improved
in the sense of (A.15b).

In the {\it massive} case we now generalize the above reasoning in (A.13).

{\bf Proposition}: {\it The adiabatic limit of Epstein and Glaser's Green's
function (A.8)}
(which exists if the normalization
constants are chosen appropriately) {\it agrees with the
infrared-improved Gell-Mann Low series in all orders}.

Proof: Let us consider the expansion
of the terms on the r.h.s. of (A.6) in normally ordered form. The terms with
vacuum
subdiagram(s) must cancel. This is due to the support property (A.7) of
$A_{0...0j_1...j_r}$.
Therefore, we may omit the terms with vacuum subdiagram(s) on the r.h.s. of
(A.6).
This omission will be denoted by a lower index $0$ in $<0\mid ...\mid 0>_0$.
(In the remark below we shall see that the adiabatic limit of an arbitrary
vacuum
diagram vanishes and, therefore, this holds true also for the terms with vacuum
subdiagram(s). However, this information is not needed in the proof here.)

The infrared-improved Gell-Mann Low expression
in n-th order is obtained by the adiabatic limit of the $I=\emptyset$ term in
(A.6)
$$\lim_{\epsilon\to 0}{1\over i^r n!}
\int d^4y_1...d^4y_n<0\mid T_{0...0j_1...j_r}(y_1,...y_n;x_1,...x_r)\mid 0>_0
g_\epsilon (y_1)...g_\epsilon (y_n)\eqno(A.16)$$
with the above mentioned omission. Consequently, we merely have to
prove that the $I\not=\emptyset$ terms in (A.6)
vanish in the adiabatic limit,
$$\lim_{\epsilon\to 0}\int d^4y_1...d^4y_n<0\mid \bar T_{0...0}(I)
T_{0...0j_1...j_r}(Y\setminus I;x_1,...x_r)\mid 0>_0
g_\epsilon (y_1)...g_\epsilon (y_n)=0\eqno(A.17)$$
for $I\not=\emptyset$. We do this by considering an arbitrary term belonging to
$<0\mid \bar T_{0...0}(I)\break T_{0...0j_1...j_r}(Y\setminus I;x_1,...x_r)\mid
0>_0$.
It has the form
$$t_1(y_1-y_s,...y_{s-1}-y_s)P^+_1(y_{j_1}-z_{i_1})...$$
$$...P^+_l(y_{j_l}-z_{i_l})
t_2(y_{s+1}-x_r,...y_n-x_r;x_1-x_r,...x_{r-1}-x_r),\eqno(A.18)$$
where $\{y_1,...y_s\}=I,\>\> \{y_{s+1},...y_n\}=Y\setminus
I,\>\>y_{j_1},...y_{j_l}\in I,\>\>
z_{i_1},...z_{i_l}\in (Y\setminus I)\cup \{x_1,...x_r\}$, and
$$P^+_j(x)=D^+_{m_j}(x),\d D^+_{m_j}(x),\d\d
D^+_{m_j}(x),...S^+_{m_j}(x),S^-_{m_j}(-x),
\>\>\>\>\>m_j>0,\>\forall y.\eqno(A.19)$$
($S^\pm_{m_j}$ means a contraction of fermionic matter fields $\psq$ and
$\psi$.) Note
that $y_{j_m}=y_{j_s}$ or $z_{i_m}=z_{i_s}$ is possible in (A.18) for $m\not=
s$.
We perform the Fourier transformation in the relative coordinates
$\tilde y_i\=d y_i-x_r\>(i=1,...,n)$ and $\tilde x_j\=d x_j-x_r\>(j=1,...,r-1)$
$$\int d^4\tilde y_1...d^4\tilde y_n\,e^{i(k_1\tilde y_1+...+k_n\tilde y_n)}
\int d^4\tilde x_1...d^4\tilde x_{r-1}\,
e^{i(p_1\tilde x_1+...+p_{r-1}\tilde x_{r-1})}(A.18)\sim$$
$$\sim\int d^4q_1...d^4q_{l-1}\,\hat t_1(...(k_i,p_j,q_m)...)\hat
P^+_1(q_1)...$$
$$...\hat P^+_{l-1}(q_{l-1})\hat P^+_l(-k_1-k_2-...-k_s-q_1-...-q_{l-1})
\hat t_2(...(k_i,p_j,q_m)...).\eqno(A.20)$$
If $t_1$ (or $t_2$) belongs to a disconnected diagram, $\hat t_1$ (rsp. $\hat
t_2$)
contains (a) $\delta^4$-distribution(s) and the number of loop integrations in
(A.20) (which seems to be $(l-1)$) is reduced. By means of (1.27) the adiabatic
limit
of (A.18) (rsp. (A.20)) can be written in the following form in momentum space
$$\sim\lim_{\epsilon\to 0}\int d^4k_1...d^4k_n\,\hat g_0(k_1)...\hat g_0(k_n)
\int d^4q_1...d^4q_{l-1}\,\hat t_1(...(\epsilon k_i,p_j,q_m)...)$$
$$\hat P^+_1(q_1)...\hat P^+_{l-1}(q_{l-1})\hat P^+_l(-\epsilon
(k_1+...+k_s)-q_1-...-q_{l-1})
\hat t_2(...(\epsilon k_i,p_j,q_m)...).\eqno(A.21)$$
Due to
$$\hat P^+_j(q)\sim\delta (q^2-m_j^2)\Theta (q^0),\>\>\>\>\>\>\forall
j=1,...l\eqno(A.22)$$
the limit (A.21) vanishes. Usually the case $l=1$ does not appear, because
there are no 1-leg
(sub)diagrams. But our proof holds true also for $l=1$. If we have no
contraction in (A.18)
($l=0$), the factor $t_1$ corresponds to a vacuum subdiagram and, therefore,
such a term is
omitted in (A.17). $\>\>\w$

Let us try to adopt this proof to the {\it massless} case. Then, the terms with
vacuum subdiagram(s) are
logarithmic divergent in the adiabatic limit (see (A.31) below). This does not
matter since these terms are omitted. Nevertheless the proof does not work:
Assuming in (A.21) for
simplicity $l=2,\>\>P^+_1,P^+_2=D^+_0$ and $\hat t_1,\hat t_2$ to be
independent on
$q\equiv q_1$, we obtain for the $q$-integration
$$\int d^4q\,\delta (q^2)\Theta (q^0)\delta ((\epsilon k-q)^2)\Theta (\epsilon
k^0-q^0)=
{\pi\over 2}\Theta (\epsilon^2 k^2)\Theta (\epsilon k^0)={\pi\over 2}\Theta
(k^2)\Theta (k^0).
\eqno(A.23)$$
The corresponding result in the massive case
($P^+_1=D^+_{m_1},\,P^+_2=D^+_{m_2}$)
is ${\pi\over 2}\Theta (\epsilon^2 k^2-(m_1+m_2)^2)\Theta (k^0)$. But in (A.23)
$\epsilon$ drops out and, consequently, the above method of proof fails.
However, the limit $\epsilon\rightarrow 0$ of (A.23) exists.

{\it Remark about the adiabatic limit of vacuum diagrams}: The adiabatic limit
can only exist if certain normalization constants are chosen in an appropriate
way. This was already noted in (3.3a), (A.15a) and it can be demonstrated in a
very simple
way for the vacuum diagrams. First we consider a {\it massive theory}. We do
not need the
explicit form of the $t$-distributions for the vacuum diagrams in this case.
Since the inner momenta go
to zero in the adiabatic limit, it suffices to know the existence of the
central
solution $\hat t_0(p_1,...p_{n-1})$ of the distribution splitting [6,8,18]
$$(D^a \hat t_0)(0,...0)=0,\>\>\>\>\>\>\forall |a|\leq \omega,\eqno(A.24)$$
and that the vacuum diagrams have the singular order (1.17)
$$\omega =4.\eqno(A.25)$$
These two assumptions are usually fulfiled in a massive, renormalizable theory
in
$d=4$ dimensions. The general splitting solution is (see (1.16))
$$\hat t(p_1,...p_{n-1})=\hat t_0(p_1,...p_{n-1})+C_0+\sum_{ij}C_{ij}p_ip_j
+\sum_{ijkl}C_{ijkl}(p_ip_j)(p_kp_l),\eqno(A.26)$$
where $C_0,\>C_{ij},\>C_{ijkl}$ are arbitrary constants.
With (1.27) we obtain for the vacuum expectation value of the n-th order
S-matrix
$$\lim_{\epsilon\to 0}\,(\Omega, S_n(g_\epsilon)\Omega)={1\over
n!}\lim_{\epsilon\to 0}
\int d^4x_1...d^4x_n\,g_\epsilon (x_1)...g_\epsilon
(x_n)t(x_1-x_n,...,x_{n-1}-x_n)=$$
$$=(2\pi)^2\lim_{\epsilon\to 0}\epsilon^{-4}\int d^4p_1...d^4p_{n-1}\,
\hat g_0(-p_1)...\hat g_0(-p_{n-1})\hat g_0(p_1+...+p_{n-1})
\hat t(\epsilon p_1,...,\epsilon p_{n-1}).\eqno(A.27)$$
Inserting (A.26) and remembering (A.24) we realize that the adiabatic limit
exists
if and only if
$$C_0=0,\>\>\>\>\>\>\>\>\>C_{ij}=0\>\>\>\>\>\forall i,j.\eqno(A.28)$$
Uniqueness of the adiabatic limit, i.e. independence on the choice of  $\hat
g_0$ in
(A.27), requires
$$C_{ijkl}=0\>\>\>\>\>\>\forall i,j,k,l.\eqno(A.28)$$
With these normalizations the vacuum diagrams vanish in the adiabatic limit
$$\lim_{\epsilon\to 0}\,(\Omega, S_n(g_\epsilon)\Omega)=0.\eqno(A.30)$$

We turn to our {\it massless Yang-Mills model} and omit the matter coupling for
simplicity.
In second order we have instead of (A.26)
$$\hat t(p)\sim (p^2)^2{\rm log}{-(p^2+i0)\over M^2}+C_0+C_2p^2,\eqno(A.31)$$
where the normalization term $\sim (p^2)^2$ is contained in the
(arbitrary) constant $M$. Inserting this into (A.27) a divergence $\sim {\rm
log}\>\epsilon$
cannot be avoided with any normalization. This is no harm for the Green's
functions
because the vacuum diagrams are absent there.
\vskip 1cm
I would like to thank Frank Krahe and Prof. G.Scharf for stimulating
discussions and for reading the
manuscript. I am grateful to Tobias Hurth for drawing my attention to some of
the
references cited below. Finally, I thank my fianc\'ee
Annemarie Schneider for bearing with me during working at this paper.
\vskip 1cm
{\trm References}
\vskip 1cm
{\obeylines
[1] M.D\"UTSCH, T.HURTH, F.KRAHE, G.SCHARF, {\it N. Cimento A} {\bf 106}
(1993), 1029
[2] M.D\"UTSCH, T.HURTH, F.KRAHE, G.SCHARF, {\it N. Cimento A} {\bf 107}
(1994), 375
[3] M.D\"UTSCH, T.HURTH, G.SCHARF, {\it N. Cimento A} {\bf 108} (1995), 679
[4] M.D\"UTSCH, T.HURTH, G.SCHARF, {\it N. Cimento A} {\bf 108} (1995), 737
[5] M.D\"UTSCH, preprint ZU-TH 10/95
[6] H.EPSTEIN, V.GLASER, {\it Ann. Inst. Poincar\'e A} {\bf 19} (1973), 211
[7] C.BECCHI, A.ROUET, R.STORA, {\it Commun. Math. Phys.} {\bf 42} (1975), 127
    C.BECCHI, A.ROUET, R.STORA, {\it Annals of Physics (N.Y.)} {\bf 98} (1976),
    287
[8] G.SCHARF, ``Finite Quantum Electrodynamics'', 2nd. ed., Springer-Verlag
(1995)
[9] M.D\"UTSCH, F.KRAHE, G.SCHARF, {\it Nuovo Cimento A} {\bf 106} (1993), 277
[10] J.C.TAYLOR, {\it Nucl.Phys. B} {\bf 33} (1971), 436
[11] A.A.SLAVNOV, {\it Theor.a.Math.Phys.} {\bf 10} (1972), 99
[12] G.'T HOOFT, {\it Nucl. Phys. B} {\bf 33} (1971), 173
[13] M.GELL-MANN, F.LOW, {\it Phys. Rev.} {\bf 84} (1951), 350
[14] M.D\"UTSCH, F.KRAHE, G.SCHARF, {\it J. Phys. G} {\bf 19} (1993) 485
[15] M.D\"UTSCH, F.KRAHE, G.SCHARF, {\it J. Phys. G} {\bf 19} (1993) 503
[16] P.PASCUAL, R.TARRACH, ``QCD: Renormalization for the Practitioner'',
Springer-Verlag (1984)
[17] T.W.CHIU, {\it Nucl.Phys. B} {\bf 181} (1981), 450
[18] M.D\"UTSCH, F.KRAHE, G.SCHARF, {\it Nuovo Cimento A} {\bf 103} (1990), 903
[19] S.K.KIM, M.BAKER, {\it Nucl.Phys. B} {\bf 164} (1980), 152
[20] N.N.BOGOLIUBOV, D.V.SHIRKOV, ``Introduction to the Theory of Quantized
Fields'', New York (1959)
[21] M.D\"UTSCH, F.KRAHE, G.SCHARF, {\it Nuovo Cimento A} {\bf 103} (1990), 871
[22] P.BLANCHARD, R.SENEOR, {\it Ann. Inst. Poincar\'e A} {\bf 23} (1975), 147
[23] T.MUTA, ``Foundations of Quantum Chromodynamics'', World Scientific,
Singapore (1987)
}
\vskip 1cm
{\trm Figure Captions}
\vskip 1cm
Fig.1. Diagrams in $x$-space representing the term
$\sim gt^{\alpha\nu\mu}_{AF}(p_1,p_2+p_3,p_4,..,.p_{n-1})$ in (2.11). The
left-handed
diagram has an external four-gluon vertex at $x_2=x_3$. In causal perturbation
theory
this four-gluon vertex is the normalization term $-{1\over 2}\delta (x_2-x_3)$
(1.18)
of the propagator $\d^\mu\d^\nu D_F(x_2-x_3)-{1\over 2}g^{\mu\nu}\delta
(x_2-x_3)$
in the (degenerate) diagram on the r.h.s..

Fig.2. Diagramatic form of the degenerate terms contributing to the connected
3-point
Green's fuctions. The terms represented by fig.1 are excluded in fig.2, since
they are
absorbed in the non-degenerate terms $\sim\tilde t^{(n)}_{B_1B_2B_3}$.

Figs.3a,b,c,d. Diagramatic form of the degenerate terms contributing to the
connected 4-point
Green's fuctions. Note that in higher orders the $t,\,\tilde t$-distributions
contain
one-particle reducible terms. We assume
all external momenta $p_i\>(i\in\{1,2,3,4\})$ and all $(p_i+p_j)\>
(1\leq i<j\leq 4)$ to be off-shell. This means that in figs.2,3a-d all momenta
of the Feynman
propagators and all non-vanishing momenta in the arguments of the $\tilde
t^{(n-1)},
\,\tilde t^{(n-2)}$-distributions are off-shell.
\bye